\documentclass[draft]{agujournal2019} 
\usepackage{url} 
\usepackage[inkscapelatex=false]{svg}
\usepackage{lineno}
\usepackage[inline]{trackchanges} 
\usepackage{soul}
\usepackage{amsmath}
\usepackage{amssymb}
\usepackage{xcolor}
\setstcolor{blue}
\usepackage[normalem]{ulem}
\definecolor{trackblue}{RGB}{0,0,255}

\draftfalse

\journalname{JGR: Planets}

\begin{document}

%
%


\title{Comparing Monte Carlo Models of Impact Alteration of Planetary Atmospheres}
%
%




\authors{M. R. Huffman\affil{1,2}, D. A. Brain\affil{1,2}, K. N. Singer\affil{3}, J. C. Johnston\affil{1}, and C. A. Caldwell\affil{1,2}}

\affiliation{1}{Laboratory for Atmospheric and Space Physics}
\affiliation{2}{University of Colorado Boulder}
\affiliation{3}{Southwest Research Institute}

\affiliation{1}{1234 Innovation Dr, Boulder, CO 80303}
\affiliation{2}{391 UCB, 2000 Colorado Ave, Boulder, CO 80309}
\affiliation{3}{1050 Walnut St $\#$300, Boulder, CO 80302}




\correspondingauthor{M. R. Huffman}{mikayla.huffman@colorado.edu}




\newcommand\db[1]{\textcolor{red}{{\scriptsize DB:} #1}}
\newcommand\ks[1]{\textcolor{blue}{{\scriptsize KS:} #1}}

\begin{keypoints}
\item Monte Carlo runs of existing models for effects of impacts on atmospheres tend to result in net growth of between $+0.01$ and $+100$ bar

\item \textit{If} the models are correct, they indicate that impacts were an important volatile delivery mechanism to the early inner planets

\item The models we investigate require algorithmic changes to function over broad initial conditions and should be applied with caution 
\end{keypoints}

%
%

%
%


\begin{abstract}
One process that affects atmospheric surface pressure is impact bombardment. The evolution of a planet's atmosphere under impact bombardment is an open question. We use a Monte Carlo method to evolve a range (0.006 to 92.5 bar) of initial atmospheres at Mars, Earth, and Venus under bombardment of $5\times10^6$ impactors using seven individual models. Since these seven models are best suited for specific impactor size regimes, we also combine these models into a composite model and compare it to other existing composites. Alterations to the existing models are required to apply to broad initial conditions. If we use each component model for every impactor, starting from present-day atmospheric pressure, we find about two or three orders of magnitude spread in the final atmospheric pressure. Given these differences, we suggest that the use of any one model to determine atmospheric change due to impact bombardment is risky. Most models and starting parameters result in net growth between $\Delta P=+0.01$ and $+100$ bar. Our composite model shows that the atmospheres of Venus, Earth, and Mars tend to grow under bombardment, with Earth's atmosphere growing most quickly. For an early Martian ($P_0=1$ bar) and an early terrestrial (with an initial pressure of $P_0=0.25$ bar) atmosphere, both tend to grow under bombardment. The results suggested here, where the models are universally applied, suggest that impact bombardment was likely a significant source of volatiles in the early Solar System. Additional work and careful consideration of how impact events affect the evolution of planetary atmospheres is needed.
\end{abstract}

\section*{Plain Language Summary}
How much atmosphere a planet has is important for life. The rocky planets in our solar system may have had more or less atmosphere in the past. Impact events, when asteroids or comets hit the atmosphere or surface of a planet, can add or remove gases from a planet's atmosphere. We compare how impacts change planetary atmospheres according to different models. We also combine several of those models into a composite model, which only uses the models in their preferred impactor size regimes. We use a Monte Carlo method to randomly generate and simulate impacts onto each body. We change the atmosphere after each impact and then use that new atmospheric pressure as the starting pressure for the next impact. We found that, if you were to throw about $10^{20}$ kg worth of impactors at Venus, Earth, and Mars, all three atmospheres grow. We also found that Earth's atmosphere grows the fastest. We tested early versions of Mars's and Earth's atmospheres and found that both would grow due to many impact events. This means that comets and asteroids may have helped give us our atmosphere.

\section{Introduction}
\begin{figure}
\centering
\noindent\includegraphics[width=\textwidth,height=\textheight,keepaspectratio]{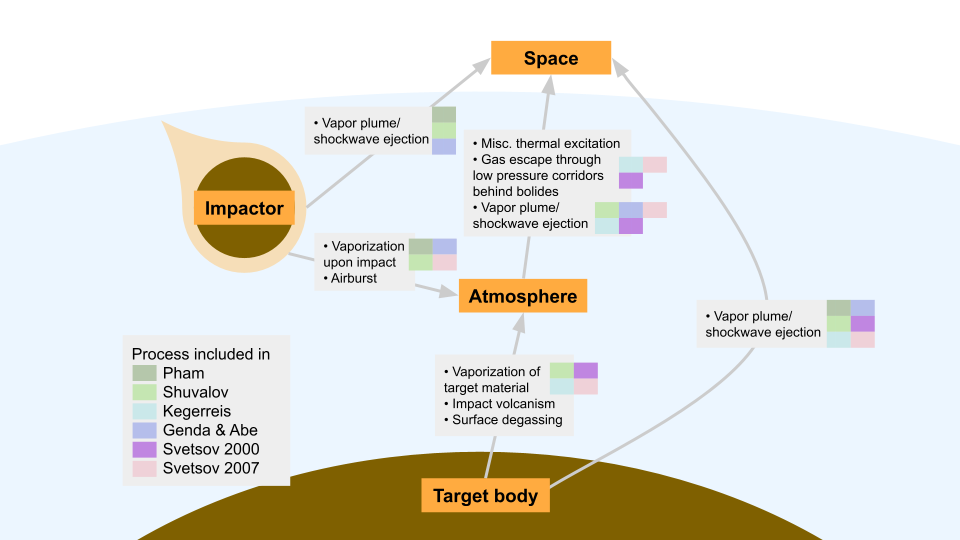}
\caption{This figure shows some of the many processes that can cause atmospheric gain or loss from an atmosphere due to an impact event. Shaded boxes represent the processes included in each individual model from the literature, discussed in Section \ref{sec:methods}. Some processes are included that are not currently described by the models listed. Vapor plumes or shockwaves can eject impactor material into space. Impactor vaporization and airbursts can transfer impactor volatiles into the target atmosphere. They can also escape via the vapor plumes/shockwaves or the low-pressure corridor that follows the bolide. Volatiles previously sequestered in the target body can be vaporized in the impact process and transferred into the atmosphere. Hypothetical impact volcanism events or later surface degassing due to residual heating can also transfer volatiles from the target body into the atmosphere on longer time scales. Vapor plumes and shockwaves can also directly transfer target body volatiles to space. Thermally excited gas, due to the impactor traveling through the atmosphere or from the cratering process, can cause atmospheric volatiles to escape to space.}
\label{fig:sourcesandsinks}
\end{figure}

Impact bombardment is an important process that affects planetary atmospheres. Unlike most evolutionary processes that act on atmospheres that only deliver volatile gases to or only remove them from an atmosphere, surface impacts and airbursts can both add and remove. \cite{brainjakosky1998,pham2009,pham2011,svetsov2007,shuvalov2009,deniem2012}. Impactors contain volatile reservoirs, a portion of which can remain trapped in a planetary atmosphere after an impact event \cite{Huffman2024}. Impact events can also remove gas from a planet’s atmosphere through several processes (figure \ref{fig:sourcesandsinks}), including thermal excitation, gas escape through low-pressure corridors trailing behind impactors, vapor plumes, and shockwave ejection. The magnitude of these changes can depend on the characteristics of the impact event: the size of the target planet, the initial atmospheric pressure and composition, the size and speed of the impactor, the impactor’s volatile content, and more.

Given that there are many thousands of impactors large enough to affect the atmospheres of planets in any given planetary system, all with their own unique characteristics, it is difficult to determine how atmospheres could respond to heavy bombardment periods. Whether atmospheres tend to shrink or grow under the net effects of many different impactors remains an open question. If atmospheres tend to shrink due to impacts, then planets in heavily bombarded systems would be less likely to harbor life. In addition, thinned atmospheres would lead to increased surface radiation, difficulty maintaining surface water, reduced greenhouse effect, larger temperature swings, and increased impact risk from smaller bolides. Alternatively, if atmospheres tend to grow due to impacts, then bombardment could reinflate planetary atmospheres that have been previously lost to space. Whether bombardment tends to grow or shrink atmospheres has ramifications for surface habitability both in our solar system and in exoplanetary systems.

Understandably, field experiments that demonstrate the effects of impact cratering on a planet’s atmosphere are difficult to perform. One field experiment was performed on the Moon, when the upper stage of an Atlas rocket was crashed into a crater at the south pole, releasing a plume that contained volatile water vapor \cite{schultz2010}. A few high-speed impact degassing experiments have been performed in the lab \cite{Gerasimov_1984,Lange_Ahrens_1982,Lange_Ahrens_1986}. These experiments provided an important basis for the models that would later provide equations for atmospheric alteration. 

There are examples of impact alteration of atmospheres outside of experimentation in our solar system. Impact flashes at Jupiter and airbursts on Earth are real-life examples of atmospheric heating due to impact events \cite{hueso2018, Brykina_2018, smith2021, vojacek2022, ozerov2024}. We also have some bounds for impact alteration from geologic history. The K-T extinction-causing impactor, for example, certainly did not blow off the entire atmosphere of Earth, given that surface life persisted after the extinction event \cite{Holroyd2014}. That being said, the Chicxulub impact did inject large amounts of vaporized rock material into the atmosphere, altering Earth's climate \cite{alvarez1980, pierazzo1998}

There are few useful experimental data, so the community tends to rely on modeling techniques. These models are optimized for various impactor size ranges. Some models are analytically derived, while others are based on shock physics hydrocode results. 

Some studies use these models for the consequences of a single impact to extrapolate to the net effects of many impacts \cite{pham2009,pham2011,deniem2012}. However, these multi-impact studies used only one model each for all types of impactors or took pieces from different models. Several models function best for only specific impactor size ranges. 

Another multi-impact study, \citeA{sinclair2020}, is mainly based on the work of \citeA{shuvalov2009} and \citeA{SCHLICHTING201581}, and incorporates some of the equations from \citeA{Kegerreis2020} and \citeA{Wyatt2019}. While both \citeA{sinclair2020} and \citeA{Wyatt2019} use a time-step method, here we use an impactor-step method, as employed in \citeA{deniem2012}. \citeA{roche2025} performed 3D smoothed particle hydrodynamics simulations for giant impacts, and their work provides equations that yield atmospheric loss. However, their model only applies to primary atmospheres, not the secondary atmospheres we investigate in this work \cite{roche2025}. In addition, they do not use initial atmospheric pressure as a free parameter \cite{roche2025}.

In this work, we will directly compare six existing models for the influence of an impact on planetary atmospheric pressure \cite{pham2011, svetsov2000, gendaabe2003, svetsov2007, shuvalov2009, Kegerreis2020}. We use an additional model, \citeA{vickerymelosh1990}, to supplement any models without gain equations. We will then combine several of them into a single composite model, such that we may compare our composite to other existing composite-type models \cite{deniem2012,SCHLICHTING201581}. For our composite, each individual model is applied only to the size range to which it applies best. Where multiple models apply to a single impact event, we average the results. By using multiple models, we can examine a wide range of physical processes accounted for in each model. We also randomly generate impactor populations using a Monte Carlo method.
The goal of our composite model is to combine the component models from the literature, applying each only within its preferred impactor size range. While this piecewise approach introduces discontinuities due to the averaging of models over different size regimes, it ensures that the models are used only in the optimal regions of the parameter space. Although our composite model does not fully capture the complex physical interactions between the impactor, surface, and atmosphere, it allows us to compare results to existing composite models, which combine these processes in different ways.

In section \ref{sec:methods} we describe our methods. Section \ref{sec:indresults} includes our data analysis and results for the individual models, while section \ref{sec:compresults} includes the analysis and results for the composite models. In section \ref{sec:casestudies}, we apply these models to the ancient atmospheres of Earth and Mars. Section \ref{sec:conclu} is our conclusions.

\section{Methods} \label{sec:methods}
In this work, we compare the results of six existing models, many of which only apply to certain impactor size regimes. We use each of these models for every impact event, even for those outside of the model's preferred size regime, as a controlled exercise to compare the models, not as a physical nor realistic assumption of universal applicability. This approach lets us isolate how each model responds to the same Monte Carlo-generated impactor population. In addition, we also compare existing composite-type models (Schlichting and de Niem), which combine the individual models in different ways. We include a novel composite model of our own as a contrast to using the individual models outside of their preferred impactor size regimes. We combine the existing component models (sector, Pham, Svetsov 2000, Genda and Abe, Svetsov 2007, Shuvalov, and Kegerreis) in their preferred size regimes, averaging between the results where there is overlap. We compare our composite model's results to other existing composite-type models and to each component model individually. The sizes, velocities, and volatile contents of the impactor populations we use are Monte-Carlo-generated. We apply the models at Venus, Earth, and Mars, with initial pressures ranging from $P_0=0.006$ bar to $P_0=92.5$ bar. 

\subsection{Existing Models} \label{subsec:eqs}
Table \ref{tab:eqns} presents the equations for gain and/or loss for each of the individual (i.e., non-composite) models we use in our composite model. It also displays the shorthand names we use for each model. Full equations and explanations for each model are given in \ref{app:eqns}. Note that we have altered these models from their original presentation to allow them to function over our broad parameter space. Our alterations are outlined in \ref{app:modelfixes}. We ran the individual models as if they applied to every impactor, not just the ones in their preferred size ranges. We also performed additional simulations for our composite model and the two existing composite-type models.

\begin{sidewaystable}
 \label{tab:eqns}
 \caption{Relevant atmospheric mass loss and gain equations for each model. Full equations are given in \ref{app:eqns}.}
 \centering
 \begin{tabular}{l l l l}
 \hline
   Model Name & Model Type & Gain Equation & Loss Equation  \\
 \hline
     Sector & Analytical & $m_\text{\text{atm,gain}}=y_\text{{imp}}m_\text{{imp}}(1-\zeta)$ & \\
     \hline
     Pham &Analytical& $m_\text{{imp}}<m_\text{{crit}} \Rightarrow m_\text{\text{atm, gain}}=m_\text{\text{imp}}y_\text{\text{imp}}f_\text{vap}$ & $m_\text{imp}<m_\text{crit} \Rightarrow m_\text{atm, loss}=0$ \\
      && $m_\text{imp}\ge m_\text{crit} \Rightarrow m_\text{atm, gain}=(1-f_\text{vel}f_\text{obl})m_\text{imp}y_\text{imp}g_\text{vap}$ & $m_\text{imp}\ge m_\text{crit} \Rightarrow m_\text{atm, loss}=m_\text{tan}f_\text{vel}f_\text{obl}$ \\
     \hline
     Svetsov 2000 &Hydrocode& & $m_\text{atm, loss}=m_\text{imp}\frac{M}{m_\text{imp}} f \left( \frac{v_\text{esc}}{v_\text{imp,0}} \left( \frac{\gamma -1}{4 \gamma} \right)^{1/2} e^{\frac{C_d}{2}\frac{M}{m_\text{imp}}} \right)$\\
     \hline
     Genda and Abe &Analytical& & $m_\text{atm,loss}=\frac{2^{2/3}Z\left( 4 \left(\frac{v_\text{imp}}{v_\text{esc}}\right)^{1-2/Z}-4^{2/Z} \right)}{3(Z-2)}\pi r_\text{imp}^2H\rho_0 \left(\frac{v_\text{imp}}{v_\text{esc}}\right)^{2/Z}$ \\
     \hline
     Svetsov 2007 &Hydrocode& $m_\text{atm,gain}=y_\text{imp}m_\text{imp}(1-\zeta(r_\text{imp}))$ & $m_\text{atm,loss}=m_\text{imp}(\psi_1+\psi_2)E$\\
     \hline
     Shuvalov &Hydrocode& $m_\text{atm, gain}=\frac{4}{3}\pi r_\text{imp}^3\rho_\text{imp}(1-\chi_\text{imp})$ & $m_\text{atm, loss}=\chi_a \frac{4}{3}\pi r_\text{imp}^3\rho_\text{imp} \frac{v_\text{imp}^2-v_\text{esc}^2}{v_\text{esc}^2}$\\
     \hline
     Kegerreis & Hydrocode &  & $m_\text{atm,loss}=0.64\left( \frac{v_\text{imp}}{v_\text{s}} \right)^2\left( \frac{m_\text{imp}}{m_\text{imp}+m_\text{target}} \right)^{0.651}f_Mm_\text{atm}$ \\
     \hline
     Schlichting &Composite& & $r_\text{imp}>r_\text{cap}\Rightarrow m_\text{atm,loss}=M_\text{cap}$ \\
      && & $r_\text{cap}>r_\text{imp}>r_\text{min}\Rightarrow m_\text{atm,loss}\approx \frac{2}{3}\pi\left( 1-\frac{r_\text{min}^2}{r_\text{imp}^2}\right)  r_\text{imp}^2 r_\text{min} \rho_\text{imp}$ \\
      && & $r_\text{imp}<r_\text{min}\Rightarrow m_\text{atm,loss}=0$ \\
      \hline
      de Niem & Composite & $m_\text{atm, gain}=(1-\zeta)y_\text{imp} m_\text{imp}$ & $m_\text{atm, loss}=\eta m_\text{imp}$ \\
 \hline
 \end{tabular}
 \end{sidewaystable}

\subsubsection{Sector Model}

The sector model is an analytic model that approximates the impact event as an expanding hemispherical vapor plume into a vacuum and assumes that the impactor material is concentrated in the outer edges of the vapor plume \cite{zeldovichraizer1967, vickerymelosh1990}. The model estimates what fraction of the liberated impactor volatiles remain gravitationally bound to the target planet. This retained fraction of the impactor's volatiles is mainly controlled by the mechanics of the expanding vapor plume. It provides equations for both atmospheric gain and loss and applies to all impactor sizes \cite{vickerymelosh1990}.

Although the sector model has equations for both gain and loss, we use only the equations for atmospheric gain, following the methodology of \citeA{deniem2012}. \citeA{deniem2012} argues that the sector model's loss equations strongly overestimate atmospheric loss for large impacts; however, the atmospheric gain equations are passable for the velocity range they use. We have a similar impactor velocity range. We employ the sector model gain equations for all models that lack an algorithm to determine loss (those being Svetsov 2000, Genda and Abe 2003, Kegerreis 2020, and Schlichting 2015). Our implementation follows the simplified volatile retention method presented by \citeA{deniem2012} that focuses on the \citeA{zeldovichraizer1967} free vapor-plume expansion solution. This version of the sector model's gain equations avoids introducing additional free parameters related to the atmospheric coupling geometry. We include it in this work as a physically motivated first order estimate of projectile volatile retention that can be used for models that do not include equations for atmospheric gain.

Given the implementation we and \citeA{deniem2012} use the retained fraction is much more dependent on the impactor velocity than on its size, and thus can be applied to a wide range of impactors as a first-order approximation for atmospheric gain. The retained fraction is nearly independent of the impactor radius. Our overall atmospheric mass gain is thus dependent on the radius of the impactor cubed, as we multiply the retained fraction by the mass of volatiles in the impactor.

\subsubsection{Tangent Plane/Pham Model}
The tangent plane model assumes that there is no atmospheric loss, only gain, below a critical impactor mass. If an impactor is more massive than the critical mass $(m_\text{crit})$, then the entire tangent plane of atmosphere above the impact site escapes \cite{MeloshVickery1989, pham2009, pham2011}. In all cases, a portion of the impactor's volatile content is added to the atmosphere \cite{MeloshVickery1989, pham2009, pham2011}. 
The tangent plane model is a computationally inexpensive, parameterized model that examines the first-order effects of impact erosion and volatile delivery due to impact events. The parameters involved are simplified versions of the full physics on display in hydrocode simulations. An important piece of the model is the impact efficiency factor, $n$. This factor determines the threshold of the model's piecewise behavior. Larger $n$ values result in less efficient erosion, and more significant volatile delivery.
The tangent plane model applies to all sizes of impactors. We use the equations from \citeA{pham2011}, which take into account impact velocity, varying impact angles, and the fraction of the impactor that is vaporized; thus, we refer to this model as the Pham model.

\subsubsection{Svetsov 2000}
\label{subsec:svet00}
The Svetsov 2000 model includes equations for atmospheric loss derived from the results of simulations in the multi-material shock physics hydrocode SOVA in 2D \cite{svetsov2000}. This model focuses on small impactors, especially bolides that interact substantially with the atmosphere, including fragmentation and deceleration during approach. This model accounts for vapor plume ejection, gas escape through low-pressure corridors, and vaporization of target material. The model thus calculates the atmospheric mass loss based on factors including the impactor's size, velocity, fragmentation, and the atmosphere of the target body. The Svetsov 2000 model is especially important for estimating erosion by smaller impactors, where atmospheric interaction, specifically, drag and fragmentation, can have significant effects on the outcome of an impact event. \citeA{svetsov2000} shows that small impactors (hundreds of meters scale) can be efficient at removing atmospheric mass, while impactors on the kilometer scale can deliver as much or more volatiles than are lost. The Svetsov 2000 model does not provide equations for atmospheric gain \cite{svetsov2000}. The Svetsov 2000 model is most applicable to impactors 0.1-1 km in diameter \cite{svetsov2000}.

\subsubsection{Genda and Abe 2003}
The Genda and Abe model is focused on the atmospheric loss processes that occur when a Mars-sized impactor impacts an Earth-sized target \cite{gendaabe2003}. Like other models, near the impact site, vapor plumes and ejecta occur which can jettison atmosphere \cite{gendaabe2003}. However, this model also focuses on atmospheric escape caused by a shock wave traveling through the target's interior, leading to ground motion far from the impact site. This process can drive atmospheric escape if the resulting surface velocity is large enough \cite{gendaabe2003}.
Genda and Abe performed spherical one-dimensional calculations of atmospheric motion for a variety of initial conditions. They found that the atmospheric loss fraction is most sensitive to the ground velocity, and does not strongly depend on the initial conditions of the target body's atmosphere. They suggest that giant impacts may not strip the entire atmosphere of a planet \cite{gendaabe2003}.
The Genda and Abe model does not provide equations for atmospheric gain \cite{gendaabe2003}. This model applies predominantly to very large (Mars-sized) impactors with diameters of about 6,000-10,000 km \cite{gendaabe2003}.

\subsubsection{Svetsov 2007}

The Svetsov 2007 model builds upon the Svetsov 2000 model \cite{svetsov2007, svetsov2000}. The Svetsov 2007 model uses the analytical equations used in the Svetsov 2000 model as a fit for 2D cylindrical geometry numerical simulation results in SOVA \cite{svetsov2007, svetsov2000}. These simulations were performed in the multi-material shock physics hydrocode SOVA \cite{svetsov2007, shuvalov1999}. The processes focused on in this model are escaping vapor plumes and wakes behind the impactor. 
Unlike Svetsov 2000, this model considers both atmospheric loss and gain. This model was developed using only vertical impacts, with the application of an enhancement factor to account for obliquity. In addition, \citeA{svetsov2007} acknowledges that the parameter space employed is narrow.
This model applies best to impactors with 0.1-10 km diameters \cite{svetsov2007}.

\subsubsection{Shuvalov 2009}
The Shuvalov model relies upon 3D numerical simulations in SOVA using multiple materials \cite{shuvalov2010abstract,shuvalov2009, shuvalov1999}. It fits a curve through numerical results for erosion efficiency to a dimensionless energy-like parameter, and then analytically builds equations for atmospheric gains and losses around that fit \cite{shuvalov2010abstract,shuvalov2009}. 
Unlike other models, Shuvalov 2009 directly simulates oblique impacts, instead of using an obliquity correction factor like \citeA{svetsov2007}. They suggest that atmospheric loss depends on the impact angle, as vertical impacts lead to the vapor plume being channeled through the impactor's wake, while oblique impacts allow the plume to develop outside of the wake, expanding more isotropically and thus affecting a larger amount of atmospheric mass \cite{shuvalov2009}. They acknowledge that their equations may not be valid for low-velocity impactors ($v_\text{imp}<10-15$ km/s) where the vapor plume is not well-developed.
The Shuvalov model best applies to impactors in the 2-24 km diameter range \cite{shuvalov2009}.

\subsubsection{Kegerreis 2020}
The Kegerreis 2020 model uses 3D smoothed particle hydrodynamics simulations to write a scaling law for atmospheric escape from low-mass atmospheres at Earth-sized planets impacted by Moon-sized impactors \cite{Kegerreis2020}. The simulations vary in impactor size, velocity, impact angles, and compositions. \citeA{Kegerreis2020} found that atmospheric loss can occur through several different mechanisms, including interaction between the impactor and target atmosphere, shock wave propagation through the planet, oscillations of the planet post-impact, and secondary impact of the bolide after initial grazing collisions. The atmospheric loss fraction is fit by a power law that depends on the impactor to total mass ratio, the impact speed relative to the escape velocity of the entire system, and impact geometry. This model is most appropriate for Mars-sized impactors hitting Earth-sized targets with small atmospheres. \citeA{Kegerreis2020} suggests that extrapolation of their equations to different impactor sizes may be difficult. This model does not include equations for atmospheric gain.

\subsection{Monte Carlo Method}
We walk through the two parts of our code, impactor generation and atmospheric evolution, in the subsections below.

\begin{table}
 \label{tab:consts}
 \caption{Constants used for each planet.}
 \centering
 \begin{tabular}{l r r r}
 \hline
   & Venus & Earth & Mars  \\
 \hline
    Planetary radius $(km)$ & 6051.8 $^{a}$& 6371 $^{b}$&3389.5 $^{c}$ \\
    Acceleration due to  & 0.00887 $^{a}$& 0.0098 $^{b}$&0.00373 $^{c}$\\
    gravity $(km/s^2)$&&&\\
    Escape velocity $(km/s)$ & 10.36 $^{a}$&11.2 $^{b}$& 5.03 $^{c}$ \\
    Scale height $(km)$ $^{\alpha}$ & 15.9 $^{a}$& 8.5 $^{b}$& 11.1 $^{c}$ \\
    Present day atmospheric &$6.5\times 10^{10}$ $^{a}$ &$1.217\times10^{9}$ $^{b}$ &$2\times10^7$ $^{c}$ \\
    surface density $(kg/km^3)$ &&&\\
    Present day atmospheric &$4.8\times10^{20}$ $^{a}$&$5.1\times10^{18}$ $^{b}$& $2.5\times10^{16}$  $^{c}$\\
    mass $(kg)$ &&&\\
    Average target crustal  &$2.8\times10^{12}$ $^{d}$ &$2.8\times10^{12}$ $^{e}$ & $2.6\times10^{12}$ $^{f}$ \\
    density $(kg/km^3)$ &&&\\
    Specific internal energy of  &0.095 $^{g}$ &0.055 $^{h}$& 0.284 $^{g}$\\
    the main atmospheric gas &&&\\
    $(km^2/s^2)$$^{\beta}$&&&\\
    Impactor size cumulative power law slope &-2.45 $^{i}$&-2.46 $^{i}$&-2.35 $^{i}$ \\
    Impactor size differential power law slope &-3.45 $^{i}$&-3.46 $^{i}$&-3.35 $^{i}$ \\
    Probability of a comet &0.3  $^{j}$&0.18 $^{j}$ & 0.06 $^{j}$ \\
 \hline
 \multicolumn{4}{l}{$^{\alpha}$We assume scale height remains constant.}\\
 \multicolumn{4}{l}{$^{\beta}$These values are given for present day atmospheric pressure.}\\
 \multicolumn{4}{l}{They are pressure dependent, and we evolve them forward at}\\
 \multicolumn{4}{l}{each impactor step.}\\
 \multicolumn{4}{l}{$^{\gamma}$These values are temperature dependent. We assume the}\\
 \multicolumn{4}{l}{temperature remains the same.}\\
 \multicolumn{4}{l}{$^{\epsilon}$Interpolated for CO$_2$ T=738.15 K.}\\
 \multicolumn{4}{l}{$^{\zeta}$Interpolated for N$_2$ T=288.15 K.}\\
 \multicolumn{4}{l}{$^{a}$NASA Venus Fact Sheet:} \\
 \multicolumn{4}{l}{ https://nssdc.gsfc.nasa.gov/planetary/factsheet/venusfact.html}\\
 \multicolumn{4}{l}{$^{b}$NASA Earth Fact Sheet: }\\
 \multicolumn{4}{l}{ https://nssdc.gsfc.nasa.gov/planetary/factsheet/earthfact.html}\\
 \multicolumn{4}{l}{$^{c}$NASA Mars Fact Sheet: }\\
 \multicolumn{4}{l}{ https://nssdc.gsfc.nasa.gov/planetary/factsheet/marsfact.html}\\
 \multicolumn{4}{l}{$^{d}$\citeA{Fegleyvenusbook}}\\
 \multicolumn{4}{l}{$^{e}$\citeA{Christensen1995}}\\
 \multicolumn{4}{l}{$^{f}$\citeA{Goossens2017}}\\
 \multicolumn{4}{l}{$^{g}$https://www.engineersedge.com/thermodynamics/thermodynamic$\_$}\\
  \multicolumn{4}{l}{properties$\_$carbon$\_$dioxide$\_$14778.htm}\\
 \multicolumn{4}{l}{$^{h}$https://www.engineersedge.com/thermodynamics/thermodynamic$\_$}\\
 \multicolumn{4}{l}{properties$\_$nitrogen$\_$table$\_$14804.htm}\\
 \multicolumn{4}{l}{$^{i}$\citeA{nesvorny2023}}\\
 \multicolumn{4}{l}{$^{j}$\citeA{olsson1987}}\\
 \end{tabular}
 \end{table}

\subsubsection{Setup and Impactor Generation}
Table \ref{tab:consts} shows the constants used for each planet. We use an asteroid density of 2700 kg m$^{-3}$ and a comet density of 1000 kg m$^{-3}$. The Svetsov 2007 model uses the atmosphere's specific internal energy. See the caption of table \ref{tab:consts} for our data sources. We determine these values for the dominant chemical species in each planetary atmosphere: CO$_2$ for Venus and Mars, and N$_2$ for Earth. As it is pressure dependent, we evolve the specific internal energy forward at each impactor step. 

After providing the necessary constants for each planet (table \ref{tab:consts}), we generate an impactor distribution of $5,000,000$ impactors. We chose this number of impactors as it results in about the correct order of magnitude for the expected impacting mass during a hypothetical heavy bombardment period in our solar system ($1.8\times10^{20}$ kg from \citeA{gomes2005}, $\sim 10^{20}$ kg from \citeA{levison2001}, and $\sim7.6\times10^{17}-6.9\times10^{20}$ kg from \citeA{chyba1990}). The median impacting mass hitting Venus per run is $(2.76_\text{-0.62}^{+2.07})\times 10^{20}$ kg, Earth $(2.71_\text{-0.61}^{+2.36})\times 10^{20}$ kg, and Mars $(7.14_\text{-2.43}^{+5.66})\times 10^{20}$ kg (higher than Venus and Earth due to its shallower power law for the impactor size-frequency distribution). These masses are equivalent to a single asteroid with a radius of $290^{+24}_{-60}$ km for Venus, $288^{+24}_{-67}$ km for Earth, and $398^{+51}_{-87}$ km for Mars.

\begin{figure}
\centering
\noindent\includegraphics[width=\linewidth]{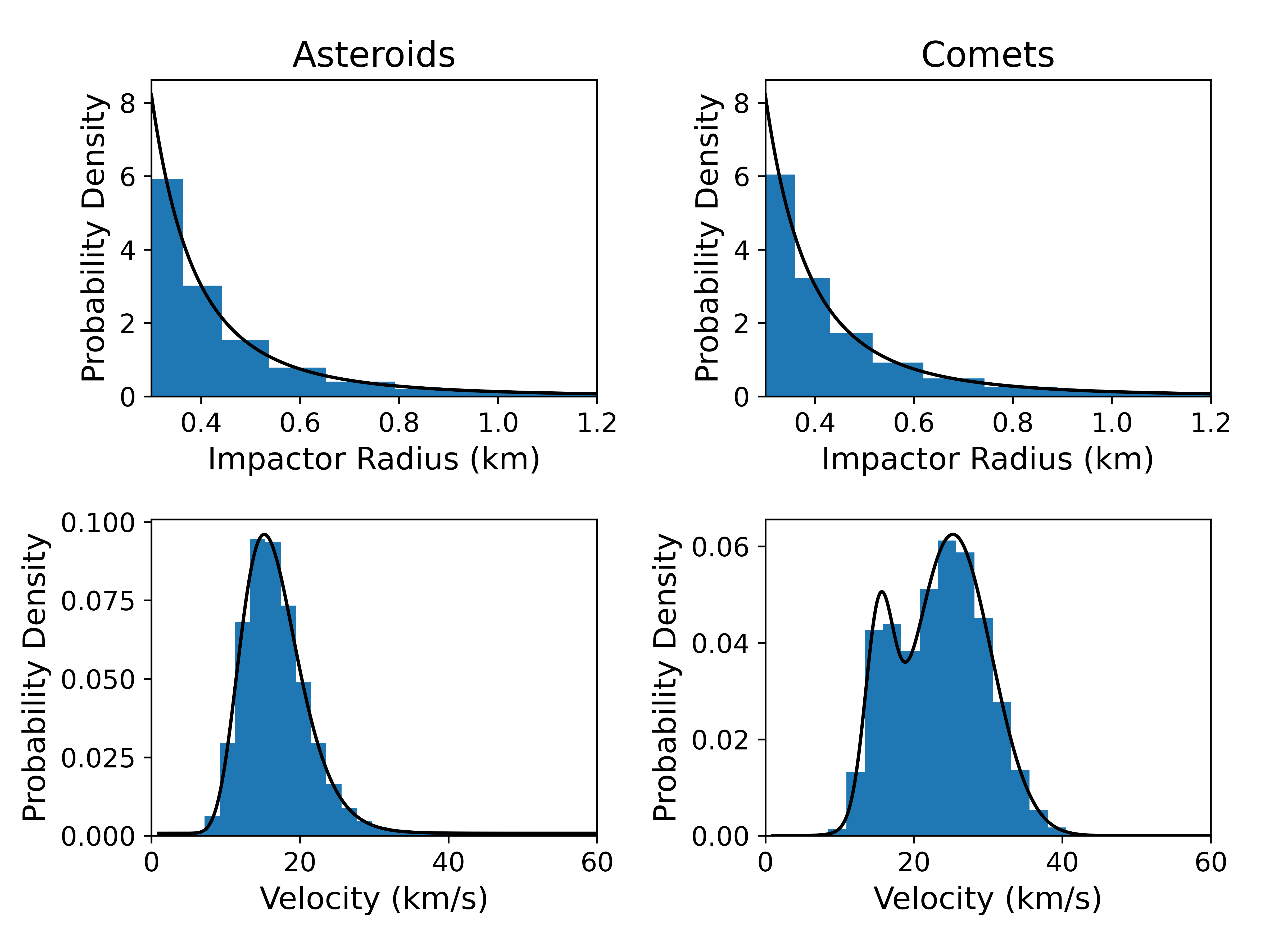}
\caption{Example probability distribution functions for impactors at Earth. The left column is asteroids, and the right column is comets. The top row shows the generated impactor sizes at Earth (blue histogram) compared to the size-frequency distribution from Figure 23 of \citeA{nesvorny2023} (black line). The bottom row shows the generated impactor velocities at Earth compared to the velocity-frequency distribution from equation 2 from \citeA{Drolshagen2020}.}
\label{fig:exPDFs}
\end{figure}

Our impactor production algorithm varies the volatile mass fraction, size, and impacting velocity of each impactor. We use a power law function for the impactor size distribution. We use the slopes from Figure 23 of \citeA{nesvorny2023} for Venus, Earth, and Mars. We artificially limit the lower and upper bounds of the impactor sizes. We only use impactors with size $r_\text{imp}=0.3-5000$ km, which is equivalent to masses $m_\text{imp}=3.1\times10^{11}-1.4\times10^{24}$ kg for asteroids or $m_\text{imp}=1.1\times10^{11}-5.2\times10^{23}$ kg for comets. If we did not provide a lower limit, we would not have the computing power to run through to the end of a bombardment period ($\approx 2\times10^{20}$ kg of impacting material). We institute an upper size limit, as giant impacts are extremely rare and likely reset the planet’s atmosphere entirely, rather than contribute to the cumulative gain or loss that we study here. For the asteroid velocities, we use equation 2 from \citeA{Drolshagen2020}. For comet velocities, we use Figure 7 from \citeA{deniem2012}. See figure \ref{fig:exPDFs} for the probability distribution functions of impactors at Earth. The distributions at Venus and Mars are visually similar.

After normalizing the probability distribution functions from the above sources, we transform them into inverse cumulative distribution functions. We then generate a random number between 0 and 1 and use the radius or velocity value associated with that number from the inverse cumulative distribution function. We also determine if the object is an asteroid or comet using probabilities for each planet given by \citeA{olsson1987}. \citeA{olsson1987} calculated impact probabilities using \citeA{Kessler1981}'s generalization of \citeA{opik1951}'s equations for two bodies in non-circular orbits. They then applied those probabilities at Venus, Earth, and Mars.

We used the same random seed for all runs. Thus, the randomly generated numbers that yield the impactor populations are the same for each set of runs. For example, the same $5,000,000$ impactors are used in the same order for the Svetsov 2000 model's first run as for the Genda and Abe model's first run. Each $n^{th}$ run across the different models applies the same impactors in the same order.

\subsubsection{Atmospheric Evolution}
Once we have our Monte Carlo-generated impactor distribution, we apply each impactor to the target planet's atmosphere. We do not account for atmospheric drag unless it is already incorporated in the model in question. We assume that the impactor's velocity at the top of the atmosphere is the same as the impacting velocity. 

\begin{figure}\centering
\noindent\includegraphics[width=1\textwidth]{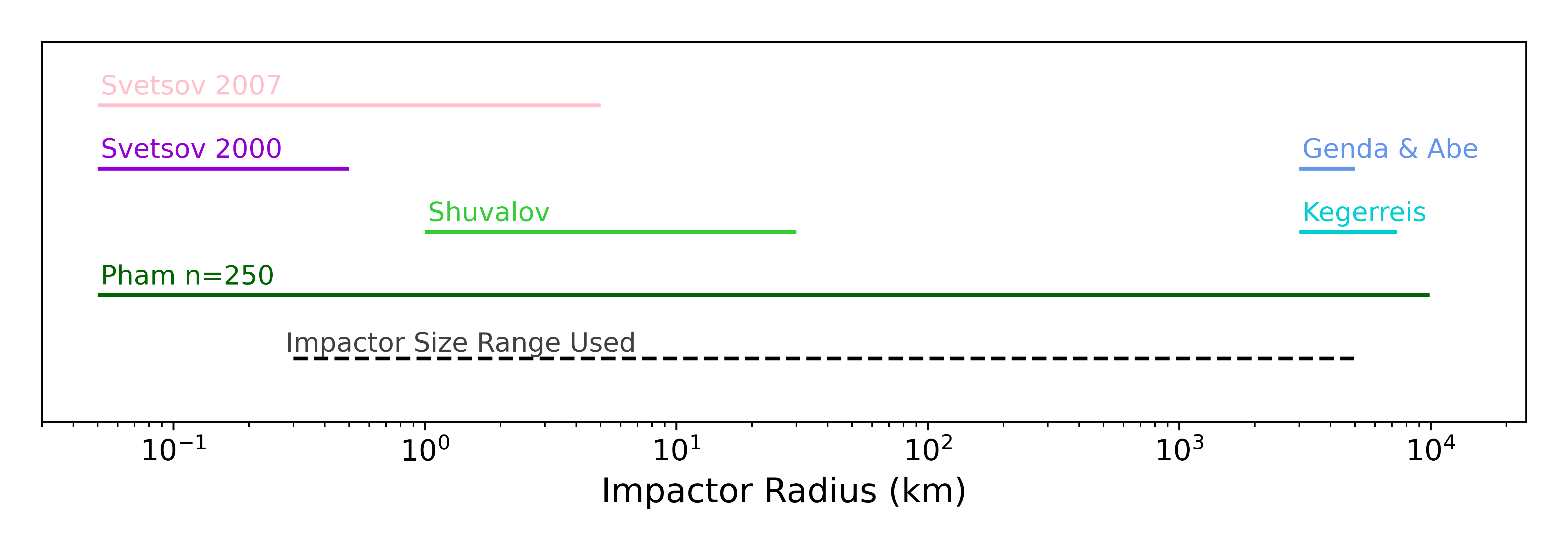}
\caption{Preferred impactor size ranges for each model. Each solid line is a different model. The dashed line is the range of impactor sizes we used.}
\label{fig:recsizes}
\end{figure}

We use the relevant equations for each individual model given in table \ref{tab:eqns} and \ref{app:eqns}. We apply them to every incident impactor, including those outside of the model's preferred size range (see figure \ref{fig:recsizes}), for the sake of comparison. If a given model indicates that more atmosphere should be lost than exists, we instead lose all of the atmosphere present. We then add together the gain and loss to get an overall change in the atmospheric pressure and evolve the atmosphere forward by that amount. This step provides a new, altered atmospheric pressure, which acts as the starting pressure for the next impactor.

We performed 30 runs of $5,000,000$ impactors each at Venus, Earth, and Mars, randomly generating the impactor populations for each run. We then took the median and interquartile ranges of the 30 runs at each impactor step. The interquartile ranges provide the error bars on plots displayed in this work.

\section{Comparison of Individual Models} \label{sec:indresults}

\subsection{Individual Model Alterations} \label{subsec:individmodels}
\begin{figure}
\centering
\noindent\includegraphics[angle=270,width=0.9\textwidth]{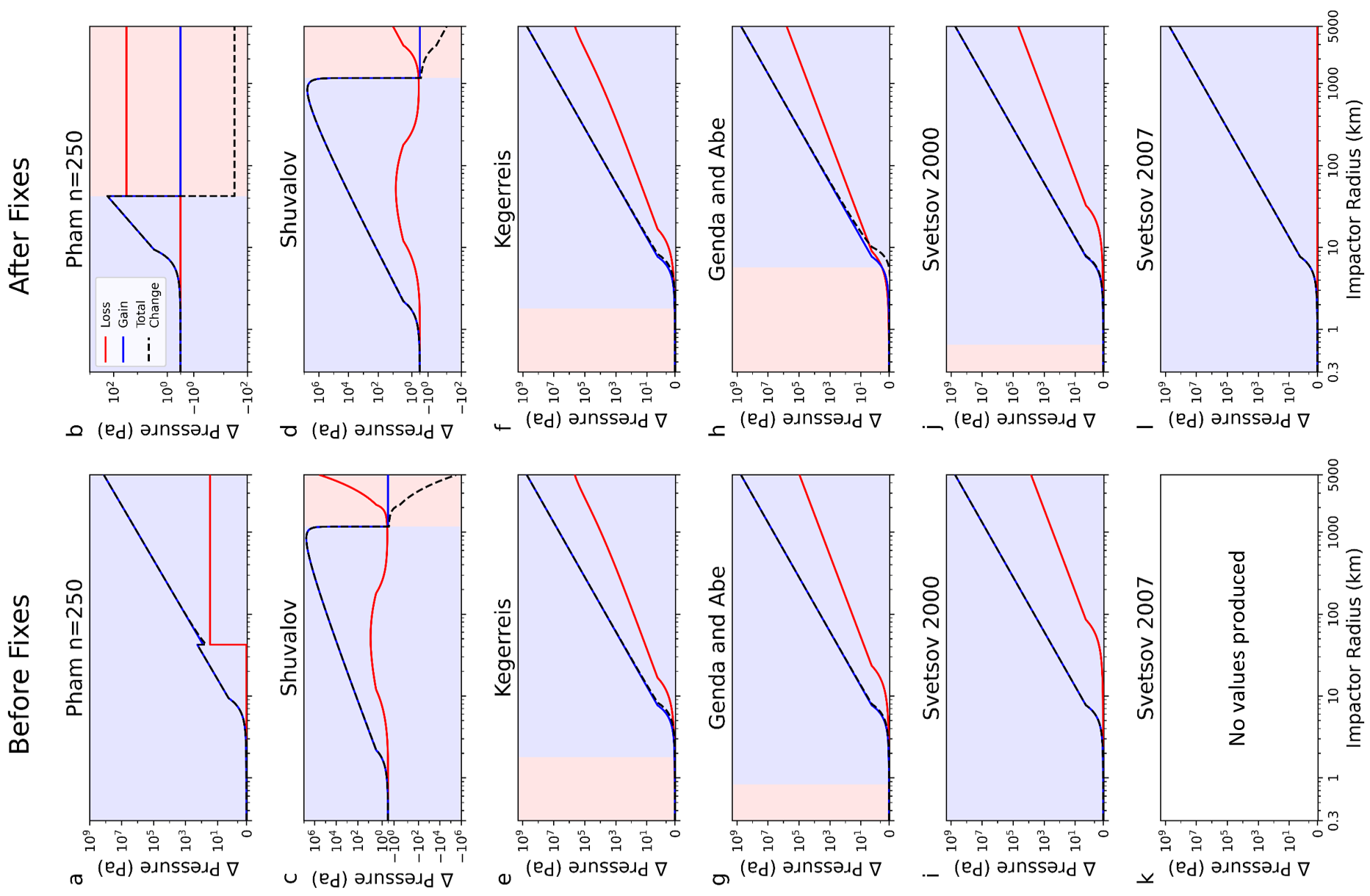}
\caption{Atmospheric pressure gain and loss due to asteroids with velocity $v_\text{imp}=16.2$ km/s (5 km/s more than the escape velocity at Earth) of a given size hitting the Earth vertically with a starting atmospheric pressure of $P_\text{0}=1$ bar. Each panel in the ``before" columns presents unaltered models from the literature, while the models in the ``after" columns are altered. See \ref{app:modelfixes} for the changes we made to each model. The blue lines are the atmospheric gain due to the asteroid impact, the red lines are atmospheric loss, and the black dashed lines are the atmospheric gain minus the atmospheric loss. The shaded regions indicate where the total change in pressure is positive (blue) or negative (red). Note that the models are dependent on more than just the impactor radius, and the full functional form of these models is more complex in the multidimensional parameter space.}
\label{fig:indgainlossbeforeafterfixes}
\end{figure}

As shown in Figure \ref{fig:indgainlossbeforeafterfixes}, each individual model predicts different amounts of atmospheric gain or loss as a function of impactor size. Note that these models depend not only on impactor size, which we have chosen to show as the independent variable in Figure \ref{fig:indgainlossbeforeafterfixes}, but also on many other factors, including initial atmospheric pressure, impactor velocity, and target body size. The models would appear even more complex and varied in the full, multidimensional parameter space. 

We made alterations to these models, as described in \ref{app:modelfixes}. We made these alterations to allow some models to produce sensible results in our parameter space. 

Following \citeA{deniem2012}, for all models that do not account for impactor angle (Svetsov 2000, Svetsov 2007, and Genda and Abe), we apply the Svetsov 2007 multiplicative factor to account for the effects of non-vertical impacts \cite{svetsov2007}. \citeA{svetsov2007} argues that an expanding vapor cloud increases with increasing impact angle. They derived the enhancement factor by integrating $ (1+2\sin(2\alpha))^2$ over an impact angle probability distribution of $\sin(2\alpha)$ from $\alpha=0-90^\circ$, yielding the final enhancement factor of $\int^{\pi/2}_0 (1+2\sin(2\alpha))^2 \sin(2\alpha) d \alpha=\frac{11}{3}+\pi$. The only non-Svetsov model we apply this enhancement factor to is Genda and Abe's model. We work from \citeA{deniem2012}'s equation 11 version of Genda and Abe's atmospheric loss derivation. This version includes approximations for nonsymmetric cases, but \citeA{deniem2012} explicitly notes that this treatment does not fully account for oblique impact geometries. We thus use the Svetsov 2007 enhancement factor only as a heuristic correction for random impactor angles. Because the Svetsov and Genda and Abe models were developed for different physical mechanisms and impactor size ranges, the validity of this correction outside the two Svetsov models remains uncertain. This enhancement factor increases the Genda and Abe losses by about 7 times. In composite-type models presented below, the enhancement factor's application to the Genda and Abe model does not affect the results significantly, as that model is only used for large, infrequent impact events.

We also alter the Pham model such that it no longer allows for gains after $m_\text{crit}$. If an impact event is energetic enough to eject the entire tangent plane of atmosphere, we argue that the volatiles sequestered in the impactor would also escape to space. Figure \ref{fig:indgainlossbeforeafterfixes} shows the models before and after our changes. Although our changes may not have appeared to significantly alter the models in this slice of the parameter space, the full model across all parameters may be more significantly altered. 
The most significant differences caused by our alterations in this slice of the parameter space are for Pham and Svetsov 2007. The Pham model now shows net loss for impactors with radii larger than the requisite radius for the critical mass. The Svetsov 2007 model is now able to yield physical values for this starting atmospheric pressure. 

Some models show net gain, where others result in net loss, and vice versa. Examining the right column (after fixes) of figure \ref{fig:indgainlossbeforeafterfixes}, we can see that the Kegerreis, Genda and Abe, and Svetsov 2000 models show net loss for small impacts, then net gain for larger ones. The gain and loss curves increase smoothly, and, where they intersect, the change from net loss to net gain occurs. Svetsov 2007 is similar, but shows net gain for all impacts in this slice of the parameter space. The Shuvalov and Pham models are more piecewise. They show net gain for smaller impacts and net loss for larger impacts. This piecewise nature comes from their differing algorithms. Pham has an explicit piecewise shape, flipping from net loss to net gain at a calculated $m_\text{crit}$ value. The Shuvalov piecewise shape comes from limiting the relative escaping projectile mass to a maximum of 1. That value is reached for giant impacts.

\subsection{Individual Model Monte Carlo Results} \label{subsec:individmodelresults}
\begin{figure}
\centering
\noindent\includegraphics[width=\textwidth,height=\textheight,keepaspectratio]{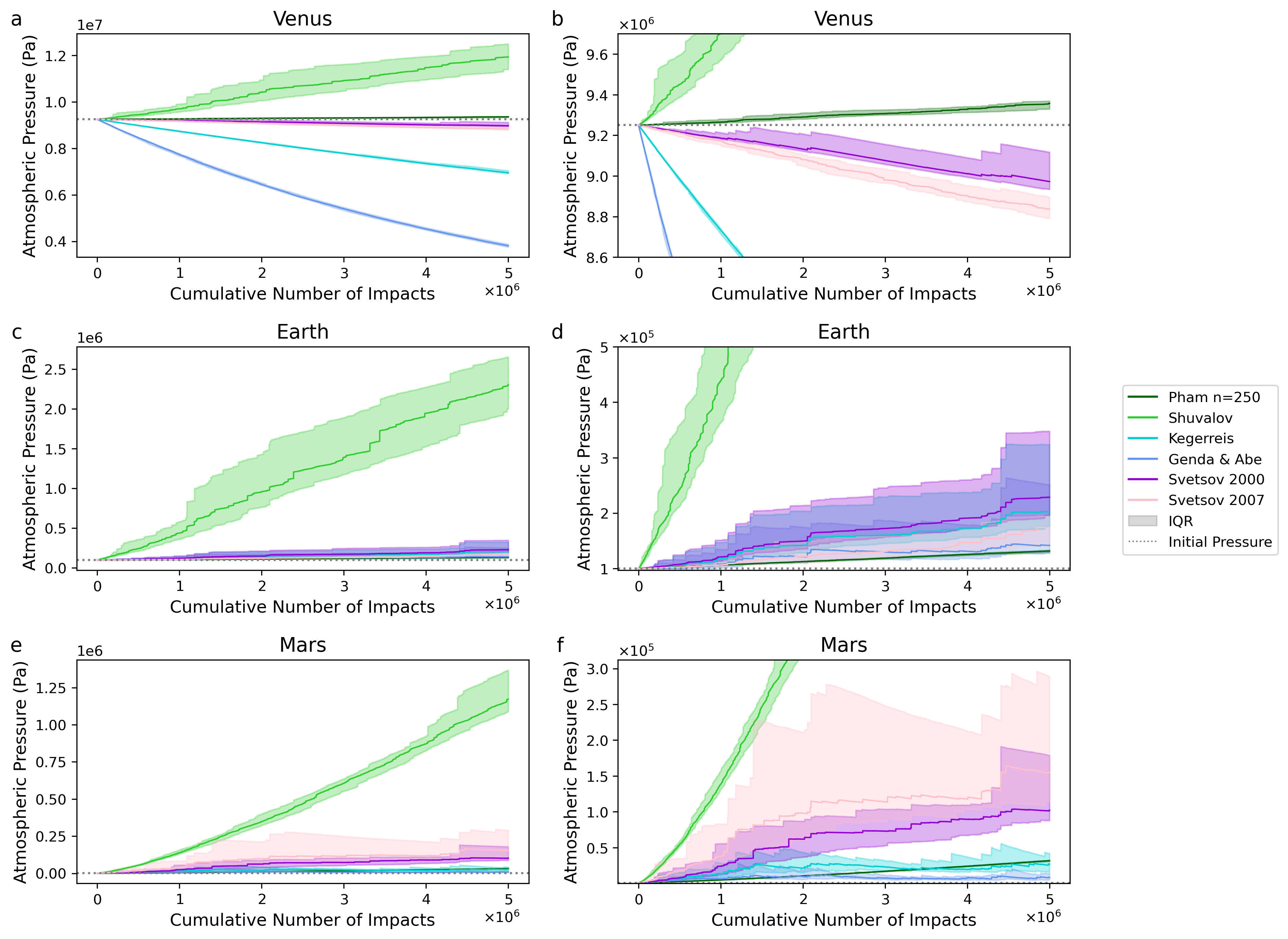}
\caption{Atmospheric change for Venus, Earth, and Mars due to each individual model. The right column shows zoomed-in portions of the vertical axis of the subplots in the left column. The horizontal axis (linear) is the number of impacts that have occurred (the number of ``impactor steps," rather than the number of timesteps). The vertical axis (logarithmic) is the atmospheric pressure after that given number of impacts has occurred. Each color is a different model. The solid line for each model is the median of 30 runs, each with different Monte Carlo-generated impactor populations. The shaded areas are the interquartile ranges (IQRs) of those 30 runs and represent the stochastic uncertainty. The horizontal dotted lines are the starting (present-day) pressures for each planet.}
\label{fig:nocomps}
\end{figure}

Figure \ref{fig:nocomps} shows the bombardment period's resultant atmospheric evolution under bombardment due to each individual model at Venus, Earth, and Mars, starting from present-day atmospheric pressure. 

At Venus, Shuvalov and Pham show net growth, while Kegerreis, Genda and Abe, and both Svetsov models show net loss. The overall spread between the final atmospheric pressure among the different models is about a factor of 3. The two Svetsov models initially give similar results, but there is no overlap of the uncertainty envelopes among the other models.
At Earth, all models result in net growth. The overall spread among the different models' final pressures is about a factor of 18. If you do not include the Shuvalov outlier, there is a factor of about 1.7 between the remaining models' final pressures. There is some overlap in the uncertainty envelopes of the models, aside from that of Shuvalov's model. 
At Mars, again, all models result in net atmospheric growth. The final spread among the results is about a factor of 140.  There is some overlap in the uncertainty envelopes of the individual models.

At Venus, some of the individual models show net growth, others show net loss. At Earth and Mars, all models show net growth. There is about a factor of 2-3 difference among the final pressures at Venus and Earth (if you do not include Shuvalov's high final pressure at Earth). At Mars, there is a factor of 140 difference between the final pressures, but we are dealing with a low-pressure atmosphere to begin with, so any small changes will be reflected greatly in a comparison like ours. Generally, Shuvalov tends to result in the highest final pressures, and Genda and Abe in the lowest.

Comparing these three planets demonstrates the sensitivity of atmospheric evolution to initial conditions and model variance. These results show that the direction of the change (net growth or net loss) is typically consistent across the different models. However, the magnitude of change varies. Note, however, that these results should be interpreted with caution, as we have been forcing these models outside of their preferred size regimes!

\section{Composite Models Comparison} \label{sec:compresults}
The differences in the final pressures outlined in section \ref{subsec:individmodelresults} may be due to applying the models out of bounds inappropriately. Composite-type models, which combine the individual models in different ways, already exist \cite{SCHLICHTING201581,deniem2012}. As another way of combining these individual models, we construct a composite model that uses the component models only in their preferred size regimes. We then compare our composite to the existing composite-type models. We do not present any one of these composites as the ``correct" or ``most physical" way of combining the individual models. The results presented here are to be used for comparison purposes, not to accurately predict the evolutionary history (or future) of planetary atmospheres. Note that we alter the de Niem model as explained in \ref{app:modelfixes}.

\subsection{Other Composite Models} \label{subsec:othercomps}
\subsubsection{de Niem 2012}
\citeA{deniem2012} is a composite model that combines analytical and hydrocode results from existing models. They use the Pham model's $m_\text{crit}$ variable to determine a threshold size. If the impactor is larger than that threshold size, they use the sector model for atmospheric gain. Otherwise, they modify the Pham model using erosion efficiencies from Svetsov 2007. The determination of atmospheric loss is similarly size-dependent. If the impactor is larger than the threshold size, they use the shock-driven free-surface motion described by Genda and Abe. For small impactors, they use the Svetsov 2000 meteor entry model with alteration by Shuvalov.

\subsubsection{Schlichting 2015}
Schlichting 2015 is a composite model that incorporates analytical methods from the sector, Pham, and Genda and Abe models \cite{vickerymelosh1990,pham2011,gendaabe2003,SCHLICHTING201581}. \citeA{SCHLICHTING201581} considers two main loss mechanisms. Giant impacts cause shockwaves to propagate through the planetary interior, which can result in loss of part or all of a planetary atmosphere, similar to the Genda and Abe model \cite{SCHLICHTING201581,gendaabe2003}. Smaller impactors can eject all the atmosphere in the tangent plane above the impact site, similar to treatment in the tangent plane model, or can eject only a fraction of the atmosphere in the tangent plane, similar to the sector model \cite{vickerymelosh1990,pham2011,SCHLICHTING201581}. These two regimes (giant impacts and smaller impactors) are analytically joined to produce a loss equation that applies to all impactor sizes. 

\subsection{Our Composite Model}
We constructed our own composite model, such that we may compare it to other existing composite-type models. For each impactor, we choose the applicable model(s) based on the impactor's size and the recommended size range of each model. Figure \ref{fig:recsizes} shows the size ranges we use for each model. Once we have decided which models to apply to the specific impactor, we use the equations for each component model from table \ref{tab:eqns} and \ref{app:eqns}. If an impactor's size falls within the preferred size ranges of multiple models, we average the atmospheric loss and the atmospheric gain due to each model. If a model requires alteration (see \ref{app:modelfixes}) to function for a particular impact event, we do not incorporate that model into the average. We are giving the individual models the best chance to succeed, only using them in their preferred size regimes, and where they do not require modification. 

We acknowledge that some of these models build upon each other. For example, the Svetsov 2007 model builds on the Svetsov 2000 model. Kegerreis uses newer hydrocode methods and could be viewed as an improvement on Genda and Abe in the giant impact regime. There are many different ways to combine these models, as shown in the variation between Schlichting and de Niem. Since others have combined the models in the ways they felt best, we present another way of combining the models, by simply averaging them.

\begin{figure}
\centering
\noindent\includegraphics[width=\textwidth]{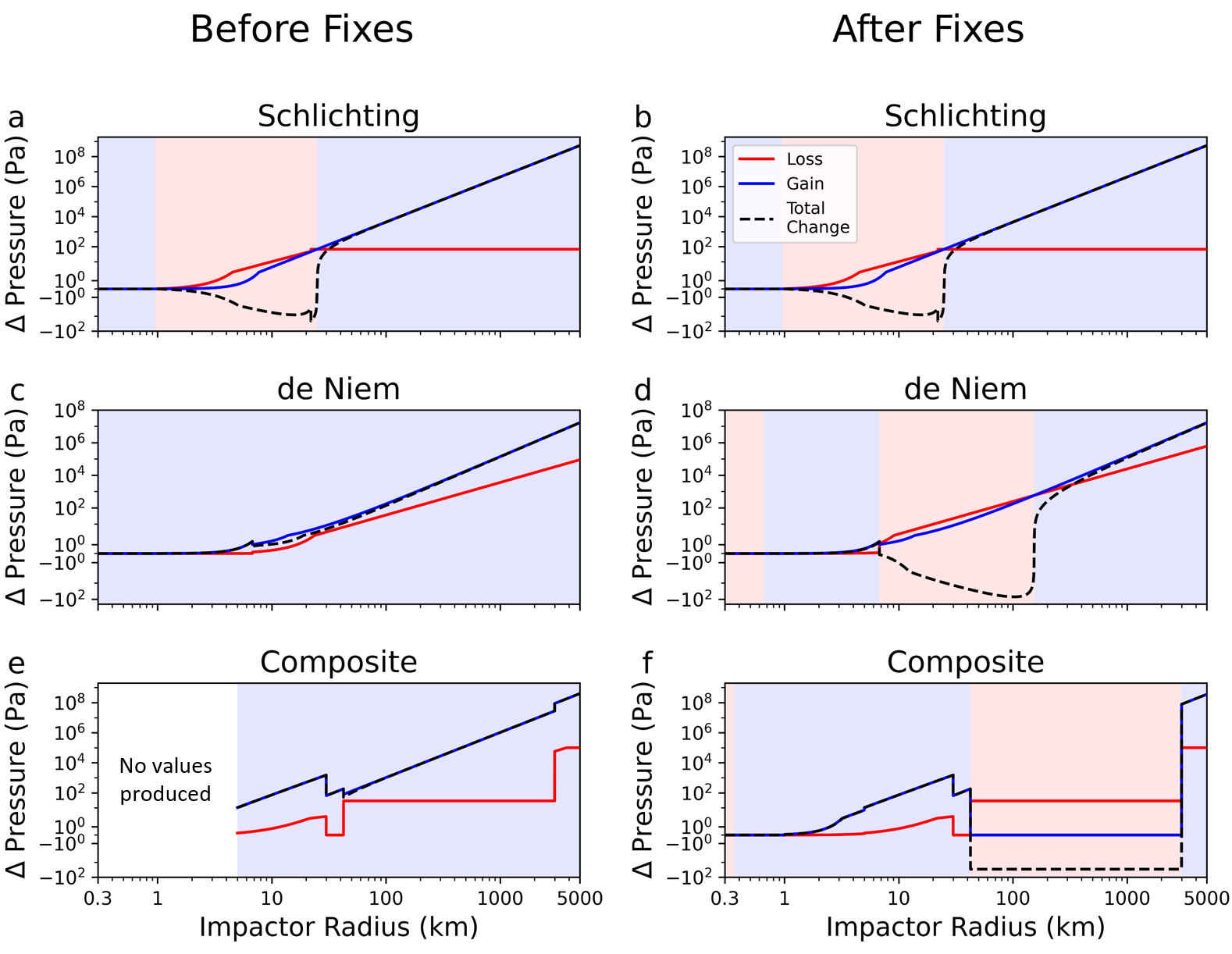}
\caption{Change in pressure due to impactor size for composite-type models, before and after our alterations. The same initial parameter set is used here as in figure \ref{fig:indgainlossbeforeafterfixes}. Axes, lines, and shaded regions have the same meanings as in figure \ref{fig:indgainlossbeforeafterfixes}.}
\label{fig:compgainlossbeforeafterfixes}
\end{figure}

Figure \ref{fig:compgainlossbeforeafterfixes} shows our composite model before (fig. \ref{fig:compgainlossbeforeafterfixes}e) and after (fig. \ref{fig:compgainlossbeforeafterfixes}f) our alterations. Recall that we have averaged together the individual models from the literature for their preferred sizes as shown in Figure \ref{fig:recsizes}. If a model requires alteration as described in \ref{app:modelfixes}, we do not include it in the average when calculating the new atmospheric pressure using our composite model. We are taking the parts of each model that are most relevant for the size of each impactor. The most significant difference caused by our alterations to the existing composite models is for the de Niem model. In this slice of the parameter space, the de Niem model's losses now outstrip its gains for small and moderately large impactors.

The resulting composite model is a piecewise function that is highly variable with impactor radius. The discontinuities occur where two size ranges and corresponding models meet. We could choose to smooth over these discontinuities, but we do not, in order to preserve the component models' distinct characteristics. Smoothing the models together would artificially blend them across the size range boundaries, possibly obscuring the behaviors of each model in its preferred size range. By retaining these discontinuities, we ensure that each component model only acts on impactors within its intended scope. However, we emphasize that the discontinuities are non-physical.

After our alterations, the model shows net gain in some regions and net loss in others, as shown in figure \ref{fig:compgainlossbeforeafterfixes}f.
Note that this model is more complex in the full parameter space, and in figure \ref{fig:compgainlossbeforeafterfixes} is only shown for varying impactor radius for asteroid impacts at present-day Earth with impactor velocity $v_\text{imp}=v_\text{esc,Earth}+5$ km/s$=16.2$ km/s. The shape of the composite model is maintained for other planets (varying surface pressure, planetary radius, etc $\--$ see table \ref{tab:consts}), though the magnitude changes.

\subsection{Composite Monte Carlo Results} \label{subsec:compresults}

\begin{figure}
\centering
\noindent\includegraphics[width=\textwidth,height=\textheight,keepaspectratio]{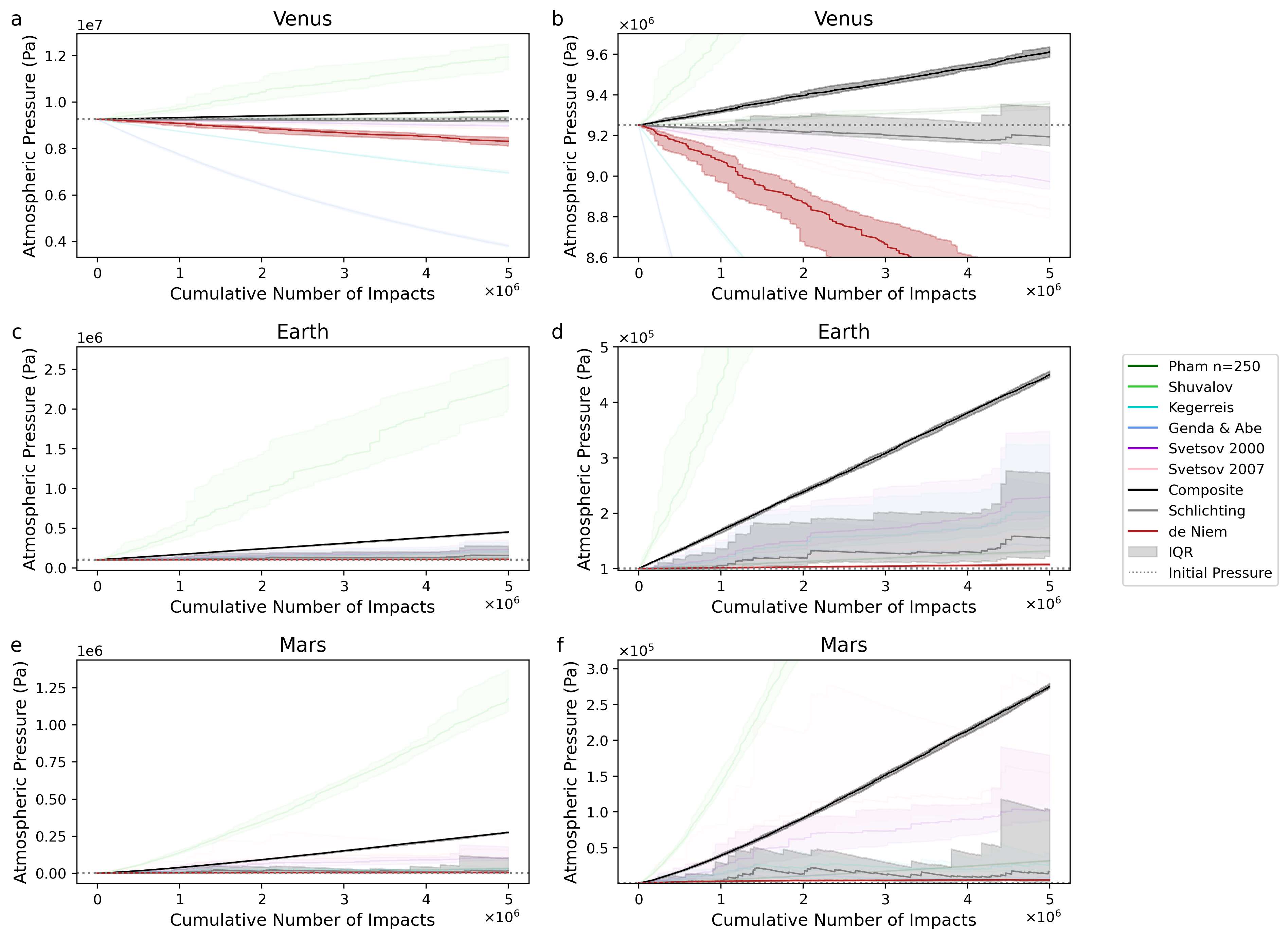}
\caption{Atmospheric change for the three planets due to each composite model. Axes, lines, and shaded regions have the same meanings as in figure \ref{fig:nocomps}. The faded colors are the component model lines shown in Figure \ref{fig:nocomps}.}
\label{fig:comps}
\end{figure}

We ran our composite model alongside the de Niem and Schlichting models. Figure \ref{fig:comps} shows our Monte-Carlo results for the composite models. 
Similar to the spread in results we see amongst the component models, the composite model results also vary. 

At Venus, our composite model results in net growth, while the de Niem and Schlichting models show net loss. There is a factor of 1.16 separating our composite and the de Niem model. 
At Earth, all three composite models result in net growth. Our composite outpaces the other two models, showing 4 times more growth than the de Niem model.  
At Mars, the results are much the same as at Earth. All three models again show net growth, with our composite model resulting in the largest growth. The final pressure for de Niem's model falls within the uncertainty envelope of Schlichting's model. There is a factor of 58 difference among the composite-type model results, again due to the low overall pressure. 

Why do these composite type models differ? Examining the right column of figure \ref{fig:compgainlossbeforeafterfixes}, we can compare which parts of the parameter space result in net gain or net loss for the different composite models for Earth with an asteroid impact, incoming at $16.2$ km/s.

The Schlichting model yields net gain for small impactors ($r_\text{imp} \lessapprox 0.96$ km), and for large impactors ($r_\text{imp} \gtrapprox 25$ km). For medium impactors, loss outstrips gain. For de Niem, however, the smallest impactors ($r_\text{imp} \gtrapprox 0.67$ km) cause net loss. Medium impactors ($0.67$ km $\lessapprox r_\text{imp} \lessapprox7$ km) cause net gain, large impactors ($7$ km $\lessapprox r_\text{imp} \lessapprox154$ km) net loss, and very large impactors ($r_\text{imp}\gtrapprox154$ km) net gain. Our composite model results in net loss for very small impactors ($r_\text{imp} \gtrapprox 0.36$ km) and large impactors ($42$ km$\lessapprox r_\text{imp} \lessapprox3000$ km). Our composite model yields net gain for medium impactors ($0.36$ km $\lessapprox r_\text{imp} \lessapprox42$ km) and giant impacts ($r_\text{imp} \gtrapprox 3000$ km).

The slopes of our impactor size frequency distribution (figure \ref{fig:exPDFs}, differential slopes of -3.45 for Venus, -3.46 for Earth, and -3.35 for Mars) results in heavy weighting toward small impactors. We see an overall trend of our composite model resulting in more gain than the other two models. Schlichting tends to appear between the de Niem model's results and ours. We will see this trend hold for varying initial pressures in section \ref{subsec:atmsize}. Comparing de Niem and Schlichting, de Niem results in loss for the smallest impactors, where Schlichting has net gain. It thus makes sense that Schlichting results in more net gain over many small cumulative impacts. Our composite model results in more net growth than both of the other composites, as the net positive change in pressure for medium-sized impactors is larger than Schlichting's the net positive change in pressure for small-sized impactors. There exists a balancing act between whether the smallest impactors result in net gain or net loss, and how large that gain or loss is for medium-sized impactors.

Recalling how the Schlichting and de Niem models combine the individual models (see section \ref{subsec:othercomps}), we can examine the physical assumptions driving these differences. Focusing on small impactors, the Schlichting model simply argues that impactors smaller than about 0.94 km (for an asteroid at Earth, see equation \ref{eq:hilkermin}) are not big enough to eject atmosphere. For small impactors, de Niem uses Svetsov’s erosion efficiency algorithm for loss, which means that their model is largely based on the assumption that small impactors break up, their diameters increasing, as they fall through an atmosphere. These differing treatments of very small impacts lead to the differences we see in the composite type models' results.

 We are left with the same issue as with the component models at Venus. Some of these composite models show net growth, others net loss. The next step is to expand our parameter space by testing other initial pressures.

\subsection{Effects of initial atmospheric pressure and planet choice} \label{subsec:atmsize}
To test how the models react to different starting pressures, we vary only the initial atmospheric pressure, holding all other parameters constant for each planet.

We ran the models at Venus, Earth, and Mars, each with initial starting pressures of 0.006 bar, 0.1 bar, 0.25 bar, 1 bar, 10 bar, and 92.5 bar. We chose 0.006, 1, and 92.5 bar as those are the present-day pressures at Mars, Earth, and Venus, respectively. We used 0.1 and 10 bar as they are nicely logarithmically spaced. We chose 0.25 bar as it is similar to the paleopressure on Earth determined by \citeA{Som2016} 2.7 billion years ago by examining vesicle sizes in basaltic lava flows. Note, again, that we are not presenting these results as true estimates of how these atmospheres would react to bombardment, but instead as another way to compare the different models.

\begin{figure}
\centering
\noindent\includegraphics[width=\textwidth,height=0.8\textheight,keepaspectratio]{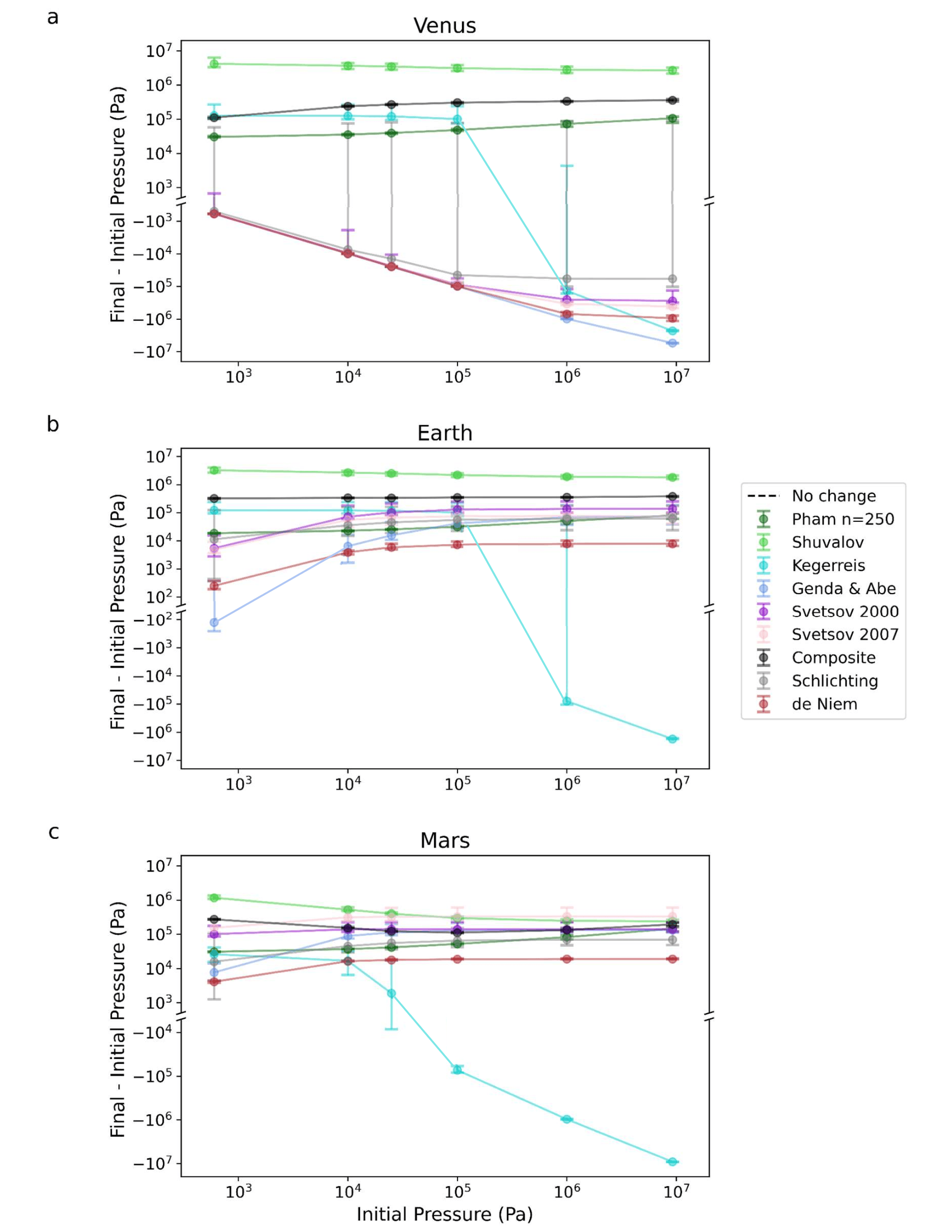}
\caption{Effects of varying the initial atmospheric pressure on the change in pressure at Venus, Earth, and Mars. Both axes are logarithmic. The horizontal axis is the starting atmospheric pressure in Pascals, and the vertical axis is the final pressure minus the initial pressure. Each color is a different model. The points are the median of 30 runs, and the error bars represent the interquartile range.}
\label{fig:changingP}
\end{figure}

Figure \ref{fig:changingP} shows the effect of initial pressure choice on the change in pressure at Venus, Earth, and Mars for all of the component and composite models. Despite the various physical processes included in each model, we observe some trends. For Earth and Mars, most models result in a net pressure increase of (very approximately) $0.01$ to $100$ bar, regardless of initial pressure. At Venus, we see two main routes for pressure change. Some models (Shuvalov, Composite, Pham) result in a net pressure change of $0.1$ to $100$ bar, but others yield a downward trend from about $-0.01$ bar for low initial pressure to $-0.1$ bar to near total loss of the atmosphere for large initial pressures (Schlichting, Svetsov 2000, Svetsov 2007, de Niem, Genda $\&$ Abe). Some models vary in whether they cause net gain or net loss, depending on the initial pressure. For example, at all three planets, the Kegerreis model results in net growth at low initial pressures, then net loss for high initial pressures. 

Why does Venus demonstrate these two routes, while the other planets do not? First, we must discuss the two structural families of models. Family A (Schlichting, Svetsov 2000, Svetsov 2007, de Niem, Genda $\&$ Abe) results in net loss at Venus. These models have loss equations that rely heavily on the atmospheric properties (density, scale height, mass, chemical composition) but have sector-style gain equations that are relatively modest. Family A results in large amounts of erosion with low volatile delivery. The models in family B (Shuvalov, Composite, Pham) are the opposite. Because their loss is comparatively weaker than family A's, they result in high atmospheric gains and limited erosion. The gain equations across both families mainly depend on the impactor properties, and less so on the atmosphere's characteristics. However, there are some differences in how atmospheric gain is handled across the two families. The sector model assumes that the volatile mass fraction of the impactor is added to the atmosphere, minus ablation from traveling through the atmosphere \cite{vickerymelosh1990}. It does not include the amount of shocked or vaporized atmosphere that remains gravitationally bound, as the Shuvalov and Pham models take into account \cite{pham2011,shuvalov2009}. The major differences are in loss efficiency across the three planets. Because the loss equations are more sensitive to atmospheric/planetary characteristics than the gain equations, even starting the three planets at the same initial pressures can result in varied atmospheric loss. Venus's specific parameter set causes Family A's models to be highly erosive, while Family B remains gain-dominated. On Earth and Mars, Family A's erosion never outpaces gain, and the models consequently remain in net-gain territory. 

Overall, it appears that the models are more responsive to planetary parameters than to initial pressure. Impact bombardment tends to result in the addition of a relatively constant amount of atmosphere at Earth and Mars, regardless of how inflated the initial atmosphere was. 

\begin{figure}
\centering
\noindent\includegraphics[width=\textwidth,height=\textheight,keepaspectratio]{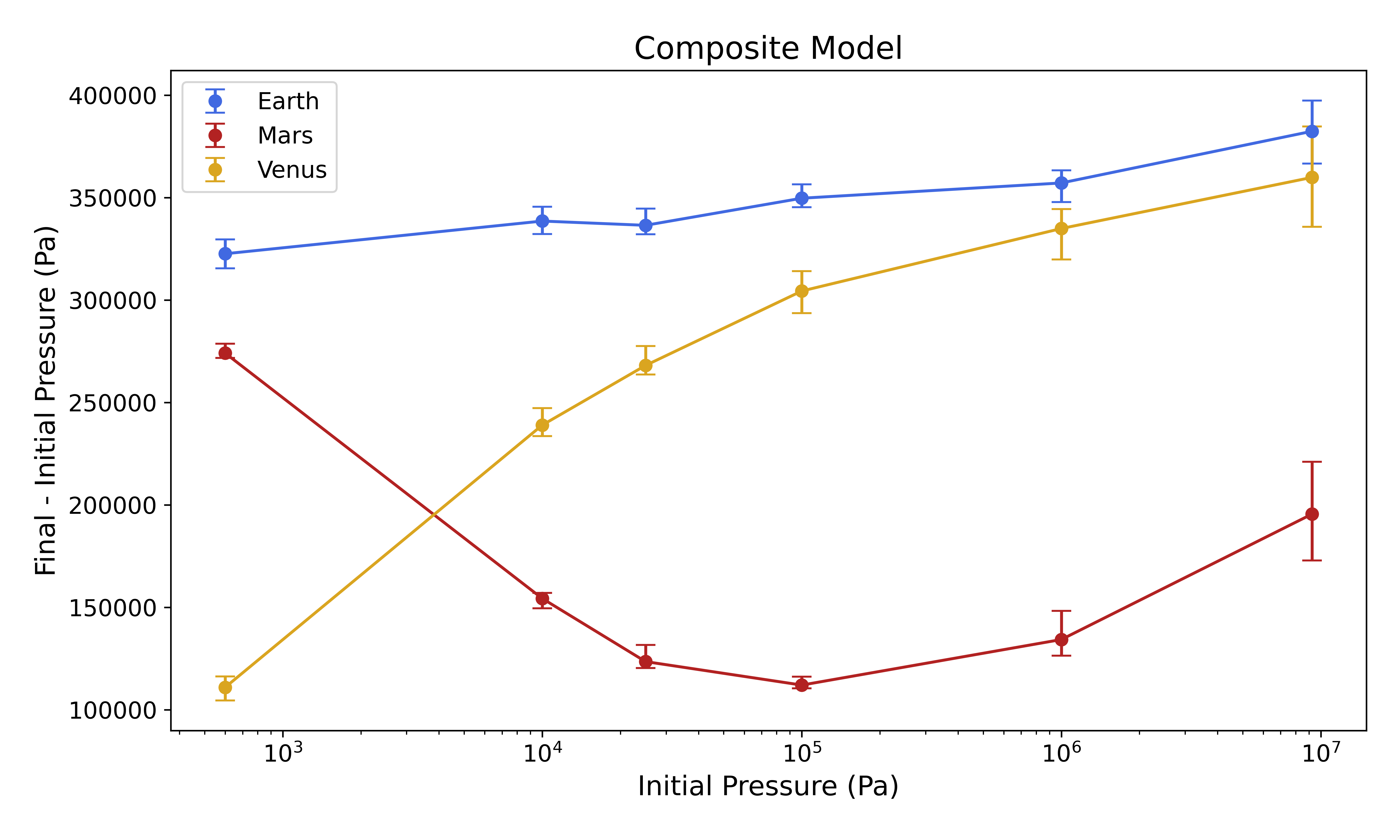}
\caption{Effects of varying the initial atmospheric pressure on change in pressure at Venus, Earth, and Mars for our composite model. The horizontal axis (logarithmic) is the starting atmospheric pressure in Pascals, and the vertical axis (linear) is the final pressure minus the initial pressure. Each color is a different planet. The points are the median of 30 runs, and the error bars represent the interquartile range.}
\label{fig:vemplot}
\end{figure}

To determine how the planetary parameters (target body size, major atmospheric constituent, power law slopes, see table \ref{tab:consts}) of Venus, Earth, and Mars affect atmospheric alteration, we ran all three planets with the same six initial atmospheric pressures. Figure \ref{fig:vemplot} shows our composite models' results at Venus, Earth, and Mars for the six starting pressures and all three planets. The curves of each planet tend to vary smoothly.

At the lowest initial pressure ($P_0=0.006$ bar), Earth's atmosphere grows the most under bombardment, and Venus's the least. At all other initial pressures tested, ($P_0=0.1$ to $P_0=92.5$ bar), Earth's atmosphere grows the most, and Mars's the least. 

Earth's atmospheric gain remains relatively constant, with a slight upward trend from $\Delta P=+3.23$ to $+3.82$ bar as the initial pressure increases. Venus's atmospheric gain increases with increasing initial pressure, rising from $\Delta P=+1.11$ to $+3.60$ bar. Mars's atmospheric pressure is more variable, with the largest increase occurring at $P_0=0.006$ bar ($\Delta P=+2.74$ bar). The gains then decrease to $\Delta P=+1.12$ bar at $P_0=1$ bar, before increasing again to $\Delta P=+1.96$ bar at $P_0=92.5$ bar. This variability demonstrates that our composite model is sensitive to more than simply the starting atmospheric pressure. Although the composite lines in figure \ref{fig:changingP} appear nearly constant, figure \ref{fig:vemplot} shows that they vary in a statistically significant way when we zoom in on the y-axis.

\subsection{Model accuracy} \label{subsec:acc}

It is unlikely that our composite model provides an accurate estimate of how impact events alter atmospheres. Intuitively, the piecewise composite function shown in figure \ref{fig:compgainlossbeforeafterfixes} does not ``look physical" but does follow from the combination of individual models that make it up. Many processes can cause atmospheric gain or loss from an atmosphere, as shown in Figure \ref{fig:sourcesandsinks}. Different models that compose our composite model account for some of these processes, but not others. Of course, some of these processes matter more at certain impactor size ranges than others. However, some of these processes (for example, surface degassing, effects of the impactor's kinetic energy on long-term thermal escape, and full airbursts) are not accounted for by any of the individual models we use here. The models that make up the composite model do not yet entirely reflect the complex net of sources and sinks that make up the impact alteration of atmospheres. 

In addition, we had to make many alterations (\ref{app:modelfixes}) such that these models could function in our parameter space. For example, the Svetsov 2007 model has difficulty when we extend it to large atmospheric pressures (e.g., Venus and Earth). The reliability of models stretched beyond their preferred parameter spaces, even while staying within their preferred impactor size ranges, is questionable. 

When the individual models are applied in isolation, that is, they are used for every single impact, even ones outside of their preferred size ranges, there are still differences between their final atmospheric pressures, by up to about one to three orders of magnitude. Whether those disagreements result in accurate outcomes when combined in our ``best case" composite model remains unclear.

The variation in the composite model and the issues we have outlined above demonstrate that there is more work to be done to incorporate more of the processes given in Figure \ref{fig:sourcesandsinks} into the models. We also advocate for additional computational modeling in different hydrocodes, which have advanced significantly over time. Implementing better atmospheric models in these hydrocodes could help us better incorporate the heating effects of impact events in planetary atmospheres.

\section{Case studies: early Mars and early Earth}
\label{sec:casestudies}
One way to see if any of the models are more accurate would be to take a paleopressure, then evolve the atmosphere forward using the models, to see if any result in the present-day pressure. In this section, we perform that analysis for Mars and Earth, noting the significant uncertainties associated with each step of the process.

Atmospheric thickness determines the minimum size of impactors that can make it to a planetary surface.  \citeA{kite2014} examined the minimum size of preserved Martian craters interbedded with river deposits from $\sim3.6$ billion years ago. If they include rimmed circular mesas, interpreted as eroded craters, they find a best-fit paleopressure of $0.9 \pm 0.1$ bar. Here, we examine the effects of a bombardment period on Mars with a starting pressure of 1 bar in detail. Note that this $\sim3.6$ Gya time period does not perfectly coincide with the $\sim4$ Gya date of the proposed Late Heavy Bombardment, but it is one of the oldest paleopressure estimates we have. 

\begin{figure}
\centering
\noindent\includegraphics[width=\textwidth,height=\textheight,keepaspectratio]{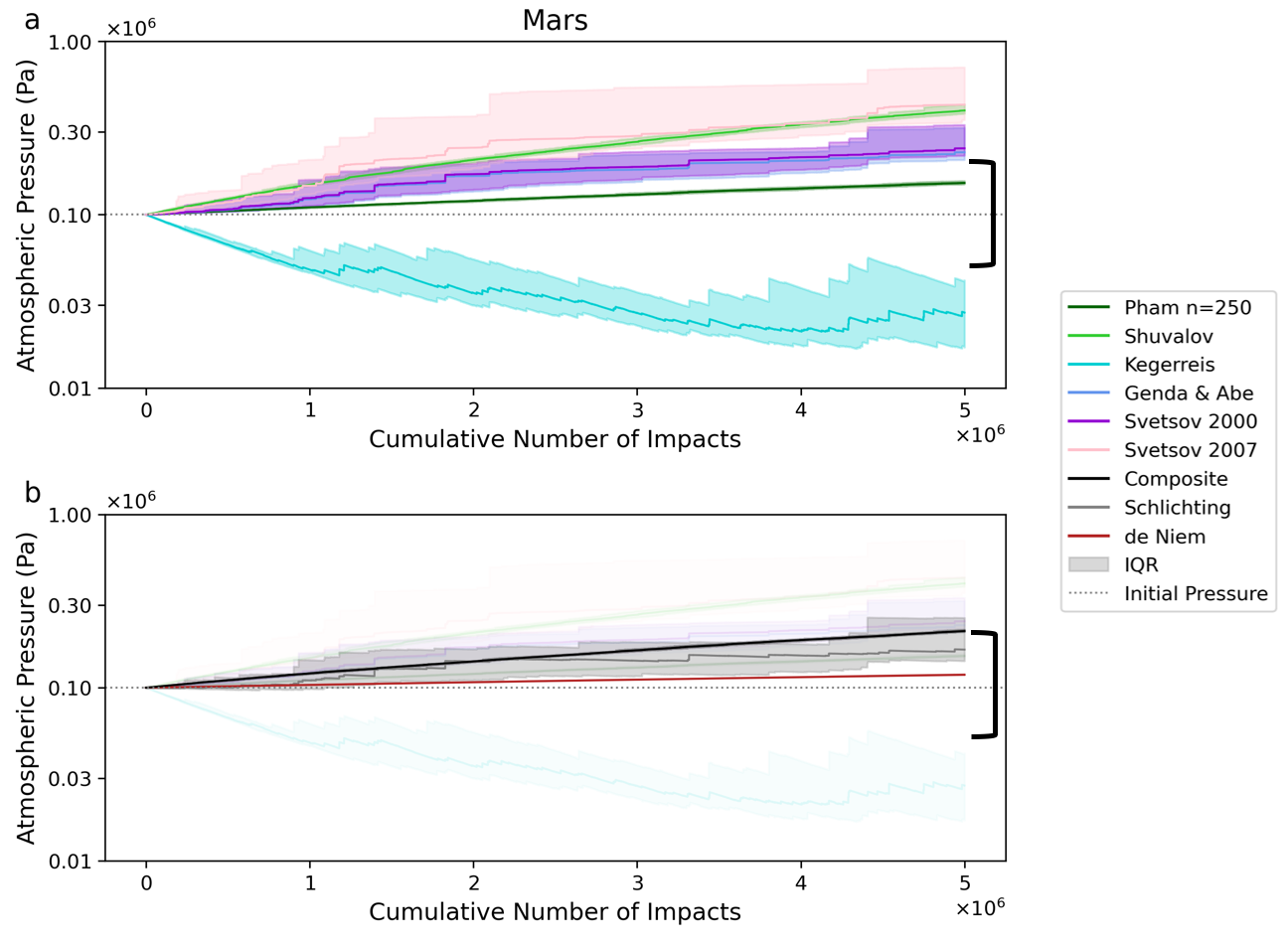}
\caption{Atmospheric change for Mars with an initial pressure of 1 bar. The axes, lines, and shaded regions have the same meanings as for figures \ref{fig:nocomps} and \ref{fig:comps}. The top panel is the individual models. The bottom panel focuses on the composite-type models, with the individual models faded to the background. The black brackets on the right side indicate the most realistic $\Delta P_\text{initial}+\Delta P_\text{bombardment}$ ranges, as discussed in the text.}
\label{fig:marscasestudy}
\end{figure}

\begin{table}[]
\centering
\caption{Final pressures from each model for Mars with an initial pressure of 1 bar and for Earth with an initial pressure of 0.25 bar.}
\label{tab:casestudies}
\begin{tabular}{llrr}
\hline
\multicolumn{1}{c}{Model Name} & \multicolumn{1}{c}{Model Type} & \multicolumn{1}{c}{\begin{tabular}[c]{@{}c@{}}Final Pressure\\ at Mars (bar)\end{tabular}} & \multicolumn{1}{c}{\begin{tabular}[c]{@{}c@{}}Final Pressure\\ at Earth (bar)\end{tabular}} \\ \hline
Pham                           & Analytical                     & $1.52^{+0.04}_{-0.02}$

                                                                     & $0.504^{+0.01}_{-0.01}$
                                                                  \\ \hline
Svetsov 2000                   & Hydrocode                      & $2.41^{+0.89}_{-0.21}$

                                                                     & $1.27^{+1.16}_{-0.18}$

                                                                      \\ \hline
Genda $\&$ Abe                 & Analytical                     & $2.29^{+0.88}_{-0.21}$

                                                                     & $0.407^{+0.305}_{-0.049}$

                                                                   \\ \hline
Svetsov 2007                   & Hydrocode                      & $4.31^{+2.78}_{-0.79}$

                                                                     & $0.916^{+0.214}_{-0.292}$

                                                                      \\ \hline
Shuvalov                       & Hydrocode                      & $3.99^{+0.34}_{-0.15}$
                                                                     & $25.4^{+3.2}_{-4.2}$

                                                                        \\ \hline
Kegerreis                      & Hydrocode                      & $0.272^{+0.145}_{-0.100}$

                            & $1.43^{+1.21}_{-0.27}$
                                                                    \\ \hline
Schlichting                    & Composite                      & $1.66^{+0.87}_{-0.23}$
                                                                   & $0.704^{+1.206}_{-0.263}$
                                                                  \\ \hline
de Niem                        & Composite                      & $1.19^{+0.01}_{-0.00}$
                                                                    & $0.309^{+0.02}_{-0.01}$
                                                                  \\ \hline
Composite                      & Composite                      & $2.12^{+0.04}_{-0.01}$
                                                                     & $3.61^{+0.09}_{-0.04}$
                                                                    
\end{tabular}
\end{table}

Figure \ref{fig:marscasestudy} demonstrates the evolution of a $P_0=1$ bar Martian atmosphere under a hypothetical bombardment period. Table \ref{tab:casestudies} presents the final pressures for each model under these initial parameters. Only one model leads to net loss. The Kegerreis model results in a final pressure of $0.272^{+0.145}_{-0.100}$ bar. All other models result in net atmospheric gain. Our composite model causes a net increase, with a final pressure of $2.12^{+0.04}_{-0.01}$ bar. According to these models, a bombardment period would likely drive atmospheric growth if it occurred at Mars with an initial pressure of 1 bar, perhaps up to $\sim10$ bar. If these model results are accurate (see section \ref{subsec:acc}), other loss processes would have to strip the additional volatiles added through impact events.   

\begin{figure}
\centering
\noindent\includegraphics[width=\textwidth,height=\textheight,keepaspectratio]{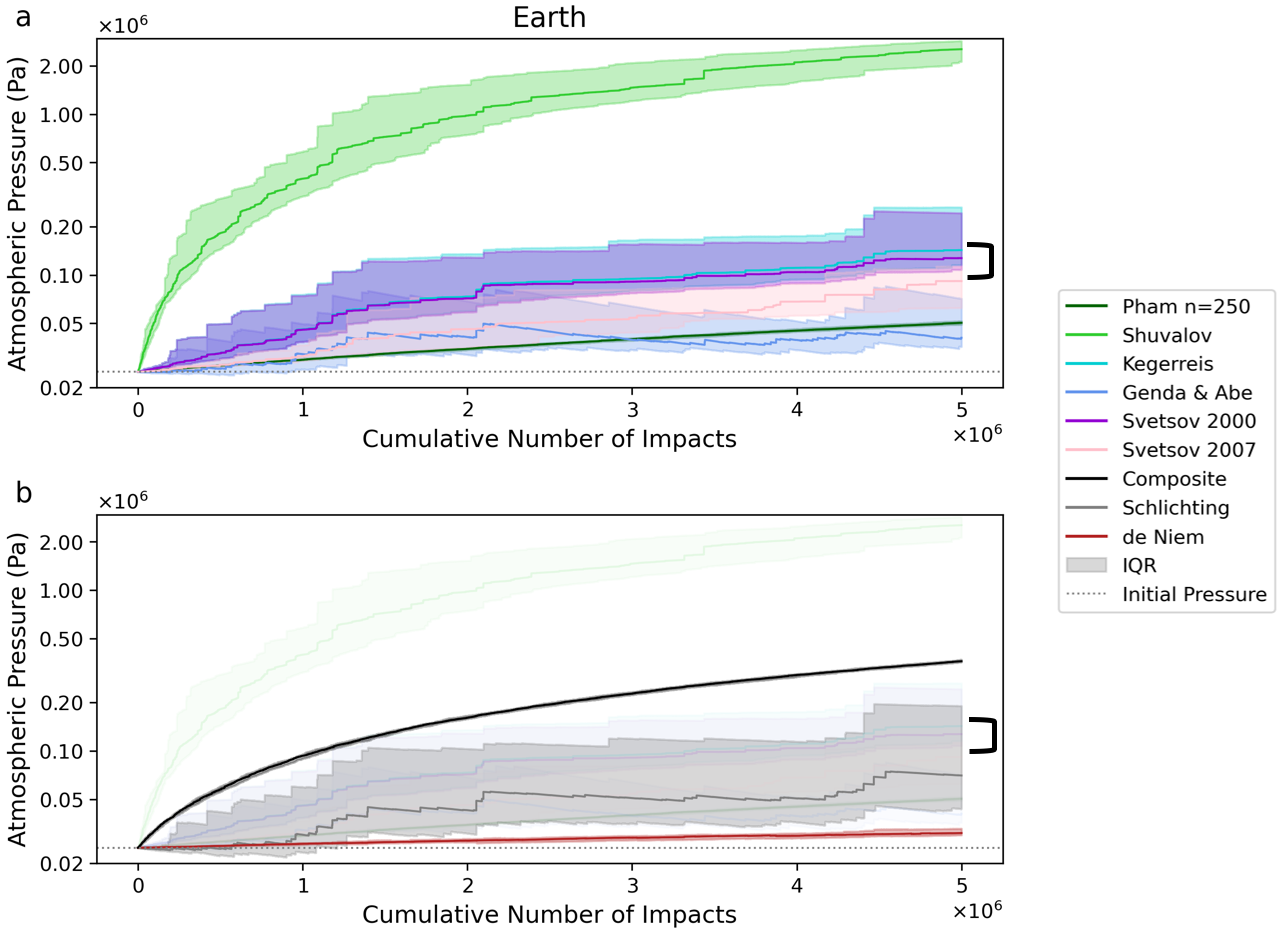}
\caption{Atmospheric change for Earth with an initial pressure of 0.25 bar. The axes, lines, and shaded regions have the same meanings as for figure \ref{fig:marscasestudy}. The black brackets on the right side indicate the most realistic $\Delta P_\text{initial}+\Delta P_\text{bombardment}$ ranges, as discussed in the text.}
\label{fig:earthcasestudy}
\end{figure}

\citeA{Som2016} suggests a terrestrial paleopressure of $0.23\pm0.23$ $(2\sigma)$ bar in the late Archaean, $\sim2.7$ Gya. They examined the size distribution of gas bubbles in basaltic lava flows using X-ray microtomography to calculate this pressure. Although the lava flows used solidified $\sim1.3$ billion years after the proposed Late Heavy Bombardment, this is still a good measure of early Earth's atmosphere.

Figure \ref{fig:earthcasestudy} shows the evolution of a $P_0=0.25$ bar terrestrial atmosphere due to a bombardment period. Table \ref{tab:casestudies} includes each model's final pressure. We use a predominantly $N_2$ atmosphere, not a $CO_2$ atmosphere, but figure \ref{fig:vemplot} demonstrates that there is only a $23\%$ difference between Venus ($CO_2$) and Earth's ($N_2$) final pressures at initial pressures of 0.25 bar. We find that all impact alteration models result in net atmospheric growth under these initial conditions. Most models' final pressures are clustered between $0.309^{+0.02}_{-0.01}$ bar (de Niem) and $3.61^{+0.09}_{-0.04}$ bar (our composite model). Shuvalov finishes higher at $25.4^{+3.2}_{-4.2}$ bar. If a bombardment period were to begin at Earth with an initial pressure of $0.25$ bar, the models used here all suggest a net increase in atmospheric pressure. 

We started these case studies with an initial atmosphere, then used each of the models to evolve the atmosphere forward through impact events. These simulations resulted in final atmospheric pressures that were altered only by bombardment. Let us now add atmospheric loss processes to determine if any of these final pressures can lead to modern-day pressures. 

\begin{equation}
    P_\text{modern}=P_\text{initial}+\Delta P_\text{bombardment}-\Delta P_\text{loss}
\end{equation}
Where $P_\text{modern}$ is the modern-day pressure, 0.006 bar at Mars and 1 bar at Earth. $P_\text{initial}$ is the initial pressure for our case studies, 1 bar for Mars and 0.25 bar for Earth. $\Delta P_\text{bombardment}$ is the change in pressure due to the bombardment period, and can be positive or negative. $\Delta P_\text{loss}$ is the corresponding change in pressure due to atmosphere being lost to space. Table \ref{tab:casestudies} displays the value $P_\text{initial}+P_\text{bombardment}$.

At Mars, total atmospheric loss to space estimates generally range from $P_\text{loss}=$0.5-2 bar \cite{Lillis2015,leblanc2015,Jakosky2017,jakosky2018,JAKOSKY2019}. Thus, the most realistic models should range from $P_\text{initial}+P_\text{bombardment}=P_\text{modern}+P_\text{loss}=0.006+[0.5-2]\approx[0.5-2]$ bar. The de Niem, Pham, and Schlichting models all result in final pressures in the appropriate range. Kegerreis yields a final pressure of $0.272^{+0.145}_{-0.100}$ bar, which is slightly too low to be within $1\sigma$ of the low end of our range. If atmospheric loss from Mars were water-dominated, perhaps the Kegerreis model could be accurate. The composite model could be plausible with more generous loss (final pressure of $2.12^{+0.04}_{-0.01}$ bar). All other models are too high to be plausible. 

At Earth, \citeA{Kislyakova2020} suggests that up to 0.39 bar of oxygen and 0.1 bar of nitrogen could have been lost. This is an upper bound, so the most realistic models should give $P_\text{initial}+P_\text{bombardment}=P_\text{modern}+P_\text{loss}=1+[0-0.49]\approx[1-1.5]$ bar. The Svetsov 2000, Svetsov 2007, Kegerreis, and Schlichting models all fall within this estimate range. The composite and Shuvalov models require multi-bar removal scenarios, and all other models would need additional volatile sources (e.g., mantle outgassing) to reach modern-day atmospheric pressures. 

The Schlichting model falls within the appropriate bounds for both planets. The Kegerreis model is close to the lower-end estimates at Mars, and falls within the correct final pressure for Earth. \textit{If} we take these simulations as physical, it would imply that Schlichting and Kegerreis are perhaps the most reliable. 

Overall, these results imply that impacts may have been a significant volatile delivery mechanism for early Mars and early Earth. Note that we are only examining the impact alteration process in isolation, followed by loss processes, however. The complex interplay of interactions between different gain and loss processes occurring at the same time may result in different atmospheric responses to bombardment than solely impacts in isolation, followed by loss. In addition, there is significant uncertainty in paleopressure and atmospheric-loss estimates for Earth and Mars. To narrow down the exact effects bombardment periods have on early rocky planet atmospheres, more work remains to be done. We suggest against placing too much confidence in these model results matching the paleopressures, and present it above as more of a thought experiment than reflective of reality.

\section{Conclusions} \label{sec:conclu}
In this work, we use Monte Carlo techniques to study the evolution of atmospheric surface pressure on Venus, Earth, and Mars-sized planets subject to impact bombardment. We compare the atmospheric evolution that results from assuming each of six different models for the influence of an impact on an atmosphere, and also construct a composite model that merges the different individual models. We compare our composite model to two other composite-type models.
The composite model is composed of averaging together individual models from the literature in their preferred size ranges. We advance the atmosphere forward after each Monte Carlo-generated impact to produce atmospheric evolution curves for present-day Venus, Earth, and Mars under bombardment. We also separately applied each individual model to every single impact event, even if the impactor was outside of the model's preferred impactor size range. We had to make many alterations to allow the models to function well for our parameter space. 

Our composite model demonstrates a method of combining existing component models. Using only one of these individual models for all impactor sizes produces results that can vary by several orders of magnitude (typically between $\Delta P=+0.01$ to $\Delta P=+100$ bar) depending on which model is applied. By combining the models so that each only acts in its preferred size range, we ensure each is used only where it performs best, even though this method produces discontinuities. This method allows us to compare our results to other composite-type models, like the de Niem and Schlichting models, while emphasizing areas where additional work is needed to incorporate additional physical processes.

We found that, if a bombardment period were to begin today, all models show that the atmospheres of Earth and Mars would grow under bombardment, but Venus's atmosphere could grow or shrink. There is some spread among the models of two or three orders of magnitude. When we varied the initial pressures for Venus, Earth, and Mars, we saw that Venus's atmospheric change direction was ambiguous, while Earth and Mars's atmospheres tended to grow under bombardment. Most models result in net growth between $\Delta P=+0.01$ and $+100$ bar. We also observed that Earth's atmosphere tends to grow more rapidly than those of Venus and Mars. These differences across the three planets, even with the same initial pressures, are due to some models' loss algorithms being highly planet-dependent. The loss depends heavily on the atmospheric and planetary characteristics, resulting in varied atmospheric evolution across the three planets even at the same initial pressures. With an early atmosphere of 1 bar, all but one model demonstrated net growth at Mars. For an early $N_2$, $P_0=0.25$ bar terrestrial atmosphere, all models resulted in net growth. Thus, \textit{if} these models are accurate, impact bombardment could have been an important driver of volatile delivery for the inner planets early in their atmospheric evolution.

Although these models agree to an extent, sometimes with overlapping uncertainty intervals, more work must be done to incorporate additional impact processes. Impact alteration is just one way that atmospheres can change. Incorporating impacts into complex networks of other atmospheric processes could lead to different outcomes than the ones presented here. 

\citeA{Zahnle_2017} argues that small fast impactors result in atmospheric erosion, while rare large impactors lead to volatile delivery. The present work suggests the state of the field is more complex; whether an impact results in net gain or net loss relies on more than just size and velocity, but on a complex interplay of many characteristics of the target body and impactor. \citeA{Zahnle_2017} also argues that the vast numbers of small impacts will dominate erosion; however, the work presented here suggests that most models will lead to net growth, even with many small impacts. We find stable, systematic trends toward net gain, which, if accurate, may shift the cosmic shoreline, resulting in more habitable worlds. We agree, however, with \citeA{Zahnle_2017}, that a better understanding of the impactor populations in other solar systems is essential for determining how atmospheres respond to bombardment.

Our work suggests that applying any single impact model to understand how atmospheres change due to impacts is risky. Given the discrepancies discussed here and advances in hydrocodes over the last decade or so, future efforts could focus on applying the hydrocodes to individual impacts over a large range of starting parameters in both impactor size and starting pressures, which may avoid some difficulties in applying these models to larger atmospheres. 

\section{Conflict of Interest Statement}
The authors have no conflicts of interest to disclose.

\appendix

\section{Model Equations}
\label{app:eqns}

\subsection{Sector Model}
\textbf{Gain:}

\begin{equation}
    m_\text{atm,gain}=y_\text{imp}m_\text{imp}(1-\zeta)
\end{equation}

Where $m_\text{atm,gain}$ is the atmospheric mass gained, $y_\text{imp}$ is the volatile mass fraction of the impactor, $m_\text{imp}$ is the mass of the impactor, and $\zeta$ is the ratio of the impactor mass lost to space compared to the original mass of the impactor and is given by:

\begin{equation}
    \zeta=C\int^1_x\tau^2(1-\tau^2)^kd\tau
    \label{eq:sectorzeta}
\end{equation}

Where $x=v_\text{esc}/\sqrt{\frac{(2k+5)2E_{vp}}{3M}} \le 1$, $\tau$ is an arbitrary constant of integration, and $C=2/B(3/2,k+1)$ is a normalization constant in terms of the Beta function. $k=1/(\gamma-1)$, and $\gamma=1.4$. In the expression for x, $E_{vp}\simeq M(v_\text{imp}^2/8-h_\text{vap})$ is the energy available for vapor plume expansion, where $h_\text{vap}=1.3\times10^7$ J kg$^{-1}$ is the specific enthalpy of vaporization for the impactor's material, and $M \simeq 2m_\text{imp}$ is the effective plume mass.

\subsection{Tangent Plane (Pham) Model}
\textbf{Gain:}
\begin{equation}
    m_\text{tan}=\frac{m_\text{atm}H}{2r_\text{target}}
\end{equation}

Where $m_\text{tan}$ is the mass of the tangent plane of the atmosphere above the impact point, $m_\text{atm}$ is the mass of the atmosphere, $H$ is the scale height of the target, and $r_\text{target}$ is the radius of the target planet.

\begin{equation}
    m_\text{crit}=nm_\text{tan}
\end{equation}

Where $m_\text{crit}$ is the critical mass, which is the minimum impactor mass that can eject the tangent plane of the atmosphere, and $n$ is an impact efficiency factor. If $n$ is higher, the critical mass is higher, and thus the erosion process is less effective and the atmospheric gains are higher. We use $n=250$, since that is close to the logarithmic average between the two extremes ($n=10$, $n=2400$) given for this value. The equations for atmospheric gain are given below:

\begin{equation}
    m_\text{imp}<m_\text{crit} \Rightarrow m_\text{atm, gain}=m_\text{imp}y_\text{imp}f_\text{vap}
    \label{eq:phamgain1}
\end{equation}

\begin{equation}
    m_\text{imp}\ge m_\text{crit} \Rightarrow m_\text{atm, gain}=(1-f_\text{vel}f_\text{obl})m_\text{imp}y_\text{imp}g_\text{vap}
    \label{eq:phamgain2}
\end{equation}

Where $m_\text{imp}$ is the mass of the impactor, $m_\text{atm, gain}$ is the atmospheric mass gained, $y_\text{imp}$ is the volatile content fraction of the impactor, $f_\text{vap}=0.55$ is the vaporization factor for $m_\text{imp}<m_\text{crit}$, $g_\text{vap}=0.42$ is the vaporization factor for $m_\text{imp}>m_\text{crit}$, $f_\text{obl}=2.17$ is a parameter that accounts for the effect of varying impact angle (from Svetsov 2007), and $f_\text{vel}\approx0.2$ is a parameter that accounts for the effect of impactor velocity.

\textbf{Loss:}
The equations for atmospheric loss are given below:

\begin{equation}
    m_\text{imp}<m_\text{crit} \Rightarrow m_\text{atm, loss}=0
\end{equation}

\begin{equation}
    m_\text{imp}\ge m_\text{crit} \Rightarrow m_\text{atm, loss}=m_\text{tan}f_\text{vel}f_\text{obl}
\end{equation}

Where $m_\text{atm, loss}$ is the atmospheric mass lost to space.

\subsection{Svetsov 2000}
\label{subsec:svet00}
\textbf{Loss:}
\begin{equation}
\label{eq:svet2000loss}
    m_\text{atm, loss}=m_\text{imp}\frac{M}{m_\text{imp}} f \left( \frac{v_\text{esc}}{v_\text{imp,0}} \left( \frac{\gamma -1}{4 \gamma} \right)^{1/2} e^{\frac{C_d}{2}\frac{M}{m_\text{imp}}} \right)
\end{equation}

Where $m_\text{atm, loss}$ is the atmospheric mass lost to space, $m_\text{imp}$ is the impactor mass, $M$ is the mass of the expanding layer of atmospheric gas, $f(x)$ is a function described below, $v_\text{esc}$ is the escape velocity of the target body, $v_\text{imp,0}$ is the initial velocity of the impactor at the top of the atmosphere, $\gamma=13/11$ is the adiabatic index of the atmospheric gas, and $C_d=2$ is the drag coefficient.

\begin{equation}
    \frac{M}{m_\text{imp}}=\frac{\rho_0}{\rho_\text{imp}}\left( \frac{H}{2r_\text{imp}}+\frac{2H^2\rho_0^{1/2}}{3r_\text{imp}^2\rho_\text{imp}^{1/2}}+\frac{H^3\rho_0}{r_\text{imp}^3\rho_\text{imp}} \right)
\end{equation}

Where $\rho_0$ is the near-surface atmospheric density, $H$ is the scale height of the target atmosphere, $r_\text{imp}$ is the radius of the impactor, and $\rho_\text{imp}$ is the density of the impactor. Svetsov uses cylindrical impactors where the diameter is equal to the height, but we use spherical impactors. The volume is slightly different, but the aspect ratio and cross-sectional area are the same.

\begin{equation}
\label{eq:svet2000int}
    f(x)=\frac{\int^1_x\left( 1-\tau^2 \right)^k d\tau}{\int^1_0\left( 1-\tau^2 \right)^k d\tau}
\end{equation}

Where $x$ is the argument of the function, $\tau$ is the variable of integration, and $k=\frac{3-\gamma}{2\gamma-2}=5$ for $\gamma=13/11$. 

We multiply the final atmospheric mass loss by an enhancement factor to account for oblique impacts as suggested by \citeA{deniem2012}. This enhancement factor comes from \citeA{svetsov2007}, and is an integral over an impact angle  distribution from $\alpha=0-90^\circ$:

\begin{equation}
    \frac{11}{3}+\pi=\int^{\pi/2}_0 (1+2\sin(2\alpha))^2 \sin(2\alpha) d \alpha
\end{equation}

\subsection{Genda and Abe 2003}
See \citeA{deniem2012} for the derivation of this particular set of equations from those presented in \citeA{gendaabe2003}.

\textbf{Loss:}

\begin{equation}
    m_\text{atm,loss}=\frac{2^{2/3}Z\left( 4 \left(\frac{v_\text{imp}}{v_\text{esc}}\right)^{1-2/Z}-4^{2/Z} \right)}{3(Z-2)}\pi r_\text{imp}^2H\rho_0 \left(\frac{v_\text{imp}}{v_\text{esc}}\right)^{2/Z}
\end{equation}

Where $m_\text{atm,loss}$ is the airmass in the region occupied by the loss cone, and Z is the shock-pressure decay exponent employed by Tonks and Melosh (1992). We use a value of 1.87, preferred by \citeA{deniem2012}, although according to de Niem, the dependence of the fore-factor on Z is not very strong. $v_\text{imp}$ is the velocity of the impactor, $v_\text{esc}$ is the escape velocity of the target, $r_\text{imp}$ is the impactor radius, $H$ is the scale height of the target, and $\rho_0$ is the density of the atmosphere at the target body’s surface.

We multiply this loss by $11/3+\pi$, the enhancement factor described in section \ref{subsec:svet00} to account for oblique impacts \cite{deniem2012,svetsov2007}.

\subsection{Svetsov 2007}
\textbf{Loss:}

The dimensionless ratio of atmospheric mass lost to space to the impactor mass due to impact vapor escaping upwards through its wake is given by:

\begin{equation}
    \psi_1(r_\text{imp})=C_1\frac{\rho_0}{\rho_\text{imp}}\left( \frac{H}{r_\text{imp}}+\frac{4H^2\rho_0^{0.5}}{3r_\text{imp}^2\rho_\text{imp}^{0.5}}+\frac{2H^3\rho_0}{r_\text{imp}^3\rho_\text{imp}} \right)\times f\left( C_2 \frac{v_\text{esc}}{v_\text{imp}} e^{C_3\left( \frac{He_0^{0.5}}{2r_\text{imp}v_\text{imp}} \right)^{C_4} \left( \frac{v_\text{imp}}{v_\text{esc}} \right)^{C_5}} \right)
    \label{eq:svet2007psi1}
\end{equation}

Where $C_1=\frac{v_\text{eE}}{v_\text{esc}}$, $v_\text{eE}=11.2$ km/s is the escape velocity of Earth, $v_\text{esc}$ is the escape velocity of the target body, $\rho_0$ is the near-surface density of the atmosphere, $\rho_\text{imp}$ is the density of the impactor, $H$ is the scale height of the target body, $r_\text{imp}$ is the radius of the impactor, and $f$ is a function given by:

\begin{equation}
    f(x)=\frac{\int^1_x\left( 1-\tau^2 \right)^5 d\tau}{\int^1_0\left( 1-\tau^2 \right)^5 d\tau}
\end{equation}

Where $\tau$ is an arbitrary integration constant. $C_2=0.27\delta^{-0.21}$, where $\delta=M_a/M_\text{aE}$ is the ratio of current atmospheric mass to present-day Earth's atmospheric mass, $M_a$ is the atmospheric mass, and $M_\text{aE}$ is the mass of the present-day Earth's atmosphere. $v_\text{imp}$ is the velocity of the impactor upon entry into the atmosphere, $C_3=1.32\delta^{0.12}$, $e_0$ is the specific internal energy of the atmospheric gas, $C_4=0.27\delta^{-0.24}$, and $C_5=0.45\delta^{0.087}$.

The ratio of the lost impactor mass to its initial mass while ignoring the atmosphere is given by:

\begin{equation}
    \zeta_v(v_\text{imp})=C_6\left( \frac{v_\text{imp}}{v_\text{esc}}-C_7 \right)^{C_8}
    \label{eq:svet2007zetav}
\end{equation}

For comets,
$C_6=0.13$,
$C_7=1.3$, and
$C_8=1.55$.
For asteroids,
$C_6=0.05$,
$C_7=2$, and
$C_8=1.4$.

For large, fast impactors, the impact vapor can propagate both upwards along the wake and sideways. The ratio of lost atmospheric mass to that of the original impactor due to the additional loss of atmosphere outside the wake is given by:

\begin{equation}
    \psi_2=\frac{3\rho_0\left( 6.25r_\text{imp}^2H+7.5r_\text{imp}H^2+4.5H^3 \right)}{2\pi\rho_\text{imp}r_\text{imp}^3}\times arctan\left( \left( \zeta_v(v_i)+10^{-3}f(\frac{0.9v_\text{esc}}{v_i}) \right)\left( \frac{r_\text{imp}}{H} \right)^4\left( \frac{\rho_\text{imp}}{\rho_0} \right)^{0.5} \right)
\end{equation}

Where $v_i$ is the impactor velocity after drag, given by:

\begin{equation}
    v_i=v_\text{imp}e^{\left( -\frac{\rho_0}{\rho_\text{imp}}\left( \frac{H}{2r_\text{imp}}+\frac{2H^2}{3r_\text{imp}^2}\left(\frac{\rho_0}{\rho_\text{imp}}\right)^{0.5}+\frac{H^3\rho_0}{r_\text{imp}^3\rho_\text{imp}} \right) \right)}
\end{equation}

Thus, the atmospheric mass loss is:

\begin{equation}
    m_\text{atm,loss}=m_\text{imp}(\psi_1+\psi_2)E
\end{equation}

Where $E=11/3+\pi$ is the enhancement factor described in section \ref{subsec:svet00} to account for oblique impacts \cite{deniem2012,svetsov2007}.

\textbf{Gain:}

The amount of impactor lost to space is approximated by:

\begin{equation}
    \zeta(r_\text{imp})=\psi_1(0.35r_\text{imp})\delta^{-1}\left( \frac{C_7v_\text{esc}}{v_\text{imp}} \right)^{0.25}+\zeta_v(v_\text{imp})\frac{2}{\pi}arctan\left( (\zeta_v(v_\text{imp}))^4 \left( \frac{r_\text{imp}}{H} \right)^{1.2}\left( \frac{\rho_\text{imp}}{\rho_0} \right)^{0.5} \right)
    \label{eq:svet2007zeta}
\end{equation}

Thus, the atmospheric mass gain is:

\begin{equation}
    m_\text{atm,gain}=y_\text{imp}m_\text{imp}(1-\zeta(r_\text{imp}))
    \label{eq:svet2007gain}
\end{equation}

\subsection{Shuvalov 2009}
\textbf{Gain:}

\begin{equation}
    \xi=\left( \frac{2r_\text{imp}}{H}\right)^3 \frac{\rho_\text{imp}}{\rho_0}\frac{v_\text{imp}^2-v_\text{esc}^2}{v_\text{esc}^2}\frac{\rho_\text{target}}{\rho_\text{target}+\rho_\text{imp}}
    \label{eq:shuxi}
\end{equation}

Where $\xi$ is a dimensionless erosional power, $r_\text{imp}$ is the impactor radius, $\rho_\text{imp}$ is the density of the impactor, $v_\text{imp}$ is the velocity of the impactor at the moment of impact, $v_\text{esc}$ is the escape velocity of the target body, $\rho_\text{target}$ is the density of the target material, $H$ is the scale height of the target body's atmosphere, and $\rho_0$ is the density of the atmosphere at the surface.

\begin{equation}
    \chi_\text{imp}=min\left( 1, \frac{0.07\rho_\text{target}v_\text{imp}(log\xi-1)}{\rho_\text{imp}v_\text{esc}} \right)
    \label{eq:chiimp}
\end{equation}

Where $\chi_\text{imp}$ is the relative escaping projectile mass.

\begin{equation}
    m_\text{atm, gain}=\frac{4}{3}\pi r_\text{imp}^3\rho_\text{imp}(1-\chi_\text{imp})
\end{equation}

Where $m_\text{atm, gain}$ is the atmospheric mass gained.

\textbf{Loss:}
\begin{equation}
    m_\text{atm, loss}=\chi_a \frac{4}{3}\pi r_\text{imp}^3\rho_\text{imp} \frac{v_\text{imp}^2-v_\text{esc}^2}{v_\text{esc}^2}
\end{equation}

Where $m_\text{atm, loss}$ is the atmospheric mass lost to space, and $\chi_a$ is a dimensionless atmospheric escape mass given as follows \cite{shuvalov2010abstract}:

\begin{equation}
    \chi_a=10^{-6.375+5.239(log\xi)-2.121(log\xi)^2+0.397(log\xi)^3-0.037(log\xi)^4+0.0013(log\xi)^5}
    \label{eq:chia}
\end{equation}

\subsection{de Niem 2012}
Note that we could not reproduce the plots given in \citeA{deniem2012} using the algorithm we pulled from their paper (below). We consistently obtained final atmospheric pressures of one or two orders of magnitude smaller than \citeA{deniem2012} reports at Mars, adopting their equations and constants. The main conclusions of our work still stand if we have interpreted or implemented these equations incorrectly. 

\textbf{Gain:}
We begin with the following equation (eq 17 in \citeA{deniem2012}):
\begin{equation}
    m_\text{atm, gain}=(1-\zeta)y_\text{imp} m_\text{imp}
\end{equation}
where $\zeta$ is the fraction of the impactor mass that is lost, and $y_\text{imp}$ is the volatile mass fraction of the impactor. The gain equation is piecewise, depending on the limiting diameter:
\begin{equation}
\label{Dlim}
    D_\text{lim}=\left(\frac{12\rho_0r_\text{tar}H^2}{\rho_\text{imp}}\right)^{\frac{1}{3}}
\end{equation}
where H is the scale height of the target atmosphere, $\rho_0$ is the density of the target atmosphere, $r_\text{tar}$ is the radius of the target body, and $\rho_\text{imp}$ is the density of the impactor.

When the impactor diameter is greater than $D_\text{lim}$, $\zeta$ is independent of the atmospheric characteristics and is given by the following equation (eq 14 in \citeA{deniem2012}):
\begin{equation}
    \zeta=C\int^1_xx^2(1-x^2)^kdx
\end{equation}
where $x=\frac{v_\text{esc}}{v_\infty}\leq 1$, $C=\frac{2}{B(3/2,k+1)}$ (where B is the beta function), $v_\infty=v_i\sqrt{4\gamma/(\gamma-1)}$, $v_i=v_\text{imp}$ exp$(\frac{-C_D}{2}\frac{m_t}{M})$ (where $v_\text{imp}$ is the meteor entry velocity and $C_D$ is the drag coefficient), $k=1/(\gamma-1)$ (where $\gamma=1.2$), and $\frac{m_t}{M}=\frac{2H}{2r_\text{imp}}+\frac{16H^2\rho_0^{1/2}}{3(2r_\text{imp})^2\rho_\text{imp}^{1/2}}+\frac{16H^3\rho_0}{(2r_\text{imp})^3\rho_\text{imp}}$.

When the impactor diameter is less than $D_\text{lim}$, 
\begin{equation}
    \zeta=\Phi\eta
\end{equation}
where $\Phi\approx\frac{1}{0.35}$. $\eta$, the erosion efficiency, is either from Svetsov or Genda and Abe's work, as shown in the loss section below \cite{svetsov2000, svetsov2007, gendaabe2003}.

\textbf{Loss:}
The atmospheric mass loss is given by
\begin{equation}
    m_\text{atm, loss}=\eta m_\text{imp}
\end{equation}
where $\eta$ is either from Svetsov (if $D<D_\text{lim}$) or Genda and Abe
(if $D>D_\text{lim}$) \cite{svetsov2000,svetsov2007,gendaabe2003}.

\subsection{Schlichting 2015}
\textbf{Loss:}
We begin with the following equation (eq 35 from \citeA{SCHLICHTING201581}):
\begin{equation}
    M_\text{cap}=2\pi\rho_\text{0} H^{2}r_\text{target}
\end{equation}
Where $M_\text{cap}$ is the mass of the atmospheric cap, $\rho_0$ is the density of the atmosphere at the surface, $H$ is the scale height, and $r_\text{target}$ is the radius of the target body.

\begin{equation} \label{eq:hilkermin}
    r_\text{min}=\left( \frac{3\rho_\text{0}}{\rho_\text{imp}}\right)^{\frac{1}{3}} H
\end{equation}
Where $r_\text{min}$ is the minimum size for an impactor to eject atmosphere, $\rho_0$ is the density of the atmosphere at the surface, $\rho_\text{imp}$ is the impactor density, and $H$ is the scale height.

\begin{equation}
    r_\text{cap}=\left( \frac{3\sqrt{2\pi}\rho_\text{0}}{4\rho_\text{imp}}\right)^{\frac{1}{3}}\sqrt{H r_\text{target}}
\end{equation}
Where $r_\text{cap}$ is the minimum size for an impactor to eject the entire cap, $\rho_0$ is the density of the atmosphere at the surface, $\rho_\text{imp}$ is the impactor density, $r_\text{target}$ is the radius of the target body, and $H$ is the scale height.

\begin{equation}
    r_\text{imp}>r_\text{cap}\Rightarrow m_\text{atm,loss}=M_\text{cap}
\end{equation}
Impactors larger than $r_\text{cap}$ eject the entire atmospheric cap.
\begin{equation}
    r_\text{cap}>r_\text{imp}>r_\text{min}\Rightarrow m_\text{atm,loss}\approx \frac{2}{3}\pi\left( 1-\frac{r_\text{min}^2}{r_\text{imp}^2}\right)  r_\text{imp}^2 r_\text{min} \rho_\text{imp} 
\end{equation}
Intermediate-sized impactors eject a fraction of the cap.
\begin{equation}
    r_\text{imp}<r_\text{min}\Rightarrow m_\text{atm,loss}=0
\end{equation}
Impactors smaller than $r_\text{min}$ do not eject atmosphere.

\subsection{Kegerreis 2020}
\textbf{Loss:}
The following equation gives the escape velocity of the system ($v_\text{s}$):
\begin{equation}
    v_\text{s}=\sqrt{\frac{2G(m_\text{target}+m_\text{imp})}{r_\text{target}+r_\text{imp}}}
\end{equation}
    where $G$ is the gravitational constant, $m_\text{target}$ is the mass of the target body, $m_\text{imp}$ is the mass of the impactor, $r_\text{target}$ is the radius of the target, and $r_\text{imp}$ is the radius of the impactor.

The overlap parameter ($d$), used to calculate the interacting mass, is given by:
\begin{equation}
    d=(r_\text{target}+r_\text{imp})(1-sin(\theta))
\end{equation}
where the impact angle $\theta=0$, in this case, means a vertical, head-on impact.

\begin{equation}
    V=\frac{4}{3}\pi r^3
\end{equation}
where $r$ is either $r_\text{imp}$ or $r_\text{target}$.

The mass is then given by:
\begin{equation}
    m=\rho V
\end{equation}
where $\rho$ is the density of the body

So, the spherical cap volume is given by:
\begin{equation}
    V_\text{cap}=\frac{\pi}{3}d^2(3r-d)
\end{equation}
where $r$ is either $r_\text{imp}$ or $r_\text{target}$.

The interacting mass fraction ($f_M$), is given by:
\begin{equation}
    f_M=\frac{\rho_\text{target}V_\text{cap,target} + \rho_\text{imp} V_\text{cap,imp}}{\rho_\text{target} V_\text{target} + \rho_\text{imp} V_\text{imp}}
\end{equation}
where $V_\text{cap,target}$ is the volume of the target's cap, $V_\text{cap,imp}$ is the volume of the impactor's cap, $V_\text{target}$ is the volume of the target body, and $V_\text{imp}$ is the volume of the impactor.

Then, the total atmospheric mass loss is given by:
\begin{equation}
    m_\text{atm,loss}=0.64\left( \frac{v_\text{imp}}{v_\text{s}} \right)^2\left( \frac{m_\text{imp}}{m_\text{imp}+m_\text{target}} \right)^{0.651}f_Mm_\text{atm}
\end{equation}

\section{Model Alterations}
\label{app:modelfixes} 

\subsection{Sector Model}

The integral that yields $\zeta$ for the sector model (equation \ref{eq:sectorzeta}) cannot function if $x=\frac{v_\text{esc}}{v_\text{1}}>1$. Thus, if $v_\text{esc}>v_\text{1}$, we set $\zeta=0$.

\subsection{Tangent Plane Model}

If the impactor mass exceeds the critical mass, we assume that there is no atmospheric mass gain, only loss. If the entire tangent plane escapes, we assume that all volatiles in the impactor escape as well.
There are no atmospheric gains if the mass of the impactor is more than the critical mass, instead of the gain equation given in equation \ref{eq:phamgain2}. Note that we include the tangent plane model in our composite model even in the region where we altered it. For all other individual models, we do not include the altered regions of the parameter space in our average.

\subsection{Svestov 2000}
When we have a small impactor and a dense atmosphere, the exponential term $e^{\frac{C_d}{2}\frac{M}{m}}$ tends to infinity (as shown below), which causes the atmospheric mass loss to also go to infinity.

\begin{equation}
    \frac{M}{m} = \frac{\rho_0 \uparrow}{\rho_m} \left( \frac{H}{2r_0 \downarrow} + \frac{2H^2 (\rho_0\uparrow)^{1/2}}{3(r_0\downarrow)^2 \rho_m^{1/2}} + \frac{H^3 (\rho_0\uparrow)}{(r_0\downarrow)^3 \rho_m} \right)
\end{equation}

As the arrows in this equation show, $\rho_0\rightarrow$ big, $r_0\rightarrow$ small, thus $\frac{M}{m}\rightarrow$ big.

In equation \ref{eq:svet2000loss},

\begin{equation}
    M_e = m\frac{M}{m} \uparrow f\left( \frac{v_e}{v_0} \left( \frac{\gamma-1}{4\gamma} \right)^{1/2} e^{\frac{C_d}{2}\frac{M}{m}\uparrow} \right)
\end{equation}

we take the exponential of $\frac{M}{m}$, which is even larger.

To more easily show how the integral step of the algorithm further increases the value, let's take the integral in equation \ref{eq:svet2000int} for $k=5$:

$$f(x)=\frac{\int^1_x\left( 1-\xi^2 \right)^5 d\xi}{\int^1_0\left( 1-\xi^2 \right)^5 d\xi}=\frac{63x^{11}-385x^9+990x^7-1386x^5+1155x^3-693x+256}{256}$$

Our exponential term then is plugged into $f(x)\propto x^{11}$, increasing the value further by raising the already large exponential term to the eleventh power. Plugging back in to equation \ref{eq:svet2000loss},

\begin{equation}
    M_e=m\frac{M}{m}\uparrow f \left( \frac{v_e}{v_0} \left( \frac{\gamma -1}{4 \gamma} \right)^{1/2} e^{\frac{C_d}{2}\frac{M}{m}} \right) \uparrow \uparrow
\end{equation}

Where ${\frac{M}{m}}{\rightarrow}$ big and ${f \left( \frac{v_e}{v_0} \left( \frac{\gamma -1}{4 \gamma} \right)^{1/2} e^{\frac{C_d}{2}\frac{M}{m}} \right)}{\rightarrow}$ very big, so $M_e$ goes to infinity when ${\rho_0}{\rightarrow}$ big and ${r_0}{\rightarrow}$ small.

The impactor's velocity can also lead to large values of atmospheric loss if $v_0\ll v_e$, as $\frac{v_e}{v_0}$ is part of the argument for $f(x)$.

So how do we fix this issue? We need to figure out where in the parameter space (of impactor size, impactor velocity, and starting atmospheric pressure) Svetsov 2000's equations yield very large atmospheric loss values and set those regions equal to zero.

\begin{figure}
    \centering
    \includegraphics[width=0.8\textwidth,height=\textheight,keepaspectratio]{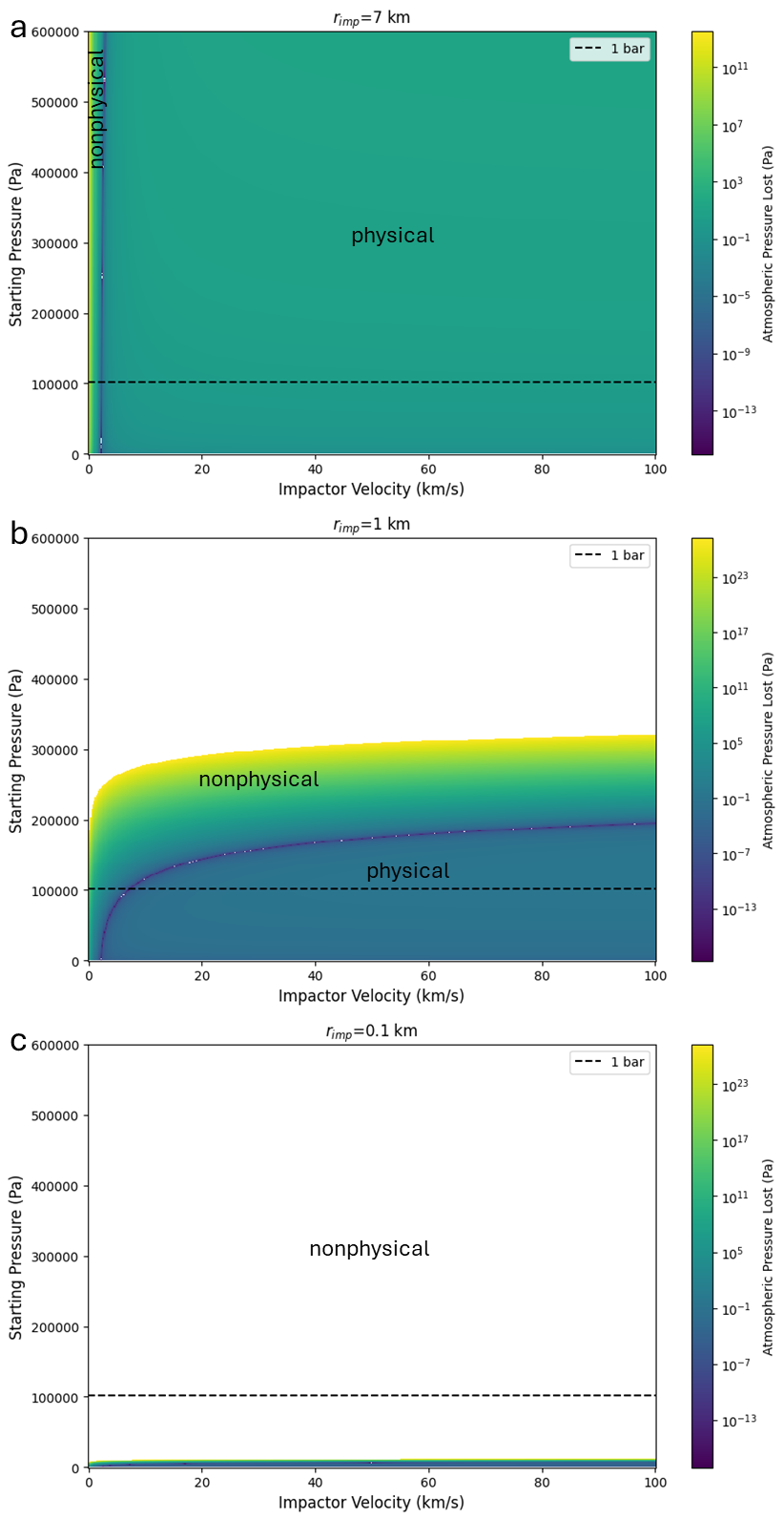}
    \caption{Plot of atmospheric pressure loss from Svetsov's 2000 model for varying impactor velocities and starting pressures for present-day Earth parameters and an impacting asteroid.}
    \label{fig:svet2000prob}
\end{figure}

We can see in Figure \ref{fig:svet2000prob}a that there is a line of local minima, to the left of which are large, nonphysical atmospheric loss values. We want to ensure those values are set to zero. As we decrease the impactor radius, that line of local minima bends over into a log function (figure \ref{fig:svet2000prob}b), and the atmospheric loss values above and to the left need to be set to zero. As the impactor radius gets smaller, the log function trends to a nearly horizontal line and the ``good region" is very small (figure \ref{fig:svet2000prob}c).

We need to determine for which impactor radius values the line of local minima goes vertical and where it goes horizontal. For comets, the line of minima in pressure-velocity space becomes vertical when $r_\text{imp}$=14 km. The vertical line is at $v_\text{imp}$=2.2 km/s. The line of minima in pressure-velocity space becomes horizontal when $r_\text{imp}$=0.3 km. For asteroids, the line of minima in pressure-velocity space becomes vertical when $r_\text{imp}$=7 km. The vertical line is at $v_\text{imp}$=2.2 km/s. The line of minima in pressure-velocity space becomes horizontal when $r_\text{imp}$=0.17 km. The above ranges are for present-day Earth parameters, but we recalculated them for each planet and starting pressure.

We need to determine whether our input parameter vector (impactor radius, impactor velocity, starting atmospheric pressure) is in the ``physical region" (to the bottom/right of the line of local minima) or in the ``nonphysical region" (to the top/left of the line of local minima). If that coordinate triplet is in the nonphysical region, we set the final atmospheric loss equal to zero. If our impactor radius is larger than the value for which the line of local minima is vertical and the impactor velocity is to the left of the line of local minima, we set the final atmospheric pressure change equal to zero. If our impactor radius is in the regime where the line of local minima is a log function, we effectively draw a cross on the plot centered at our input parameter vector. If the atmospheric pressure loss minima in the horizontal and vertical directions are below or to the right of our input vector, it means we are in the ``nonphysical region" and thus we set the final atmospheric pressure change equal to zero. If our impactor radius is smaller than the value for which the line of local minima is horizontal, we set the final atmospheric pressure change equal to zero.

\subsection{Genda and Abe 2003}
Thus, if $v_\text{imp}<v_\text{esc}$, the loss cone term $(4(v_\text{imp}/v_\text{esc})^{1-2/Z}-4^{2/Z})$ can be negative, and is no longer physical. We do not account for drag, so this domain limit does not apply to our use case.

\subsection{Svetsov 2007}

Svetsov's 2007 atmospheric gains can sometimes be negative or complex. If we have a large starting pressure, we often get negative values for the atmospheric gain.

When we have $\rho_0 \rightarrow$ large, $m_\text{atmgain} < 0$. Using equation \ref{eq:svet2007psi1}:

\begin{equation}
    \psi_1(0.35r_0) = C_1 \frac{\rho_0\uparrow}{\rho_m} \left( \frac{H}{0.35r_0}+\frac{4H^2 (\rho_0\uparrow)^{0.5}}{3(0.35r_0)^2 \rho_m^{0.5}} + \frac{2H^3 \rho_0\uparrow}{(0.35r_0)^3 \rho_m}\right) \times f\left( C_2 \frac{v_e}{v} e^{C_3 \left( \frac{He_0^{0.5}}{2(0.35r_0) v} \right)^{C_4} \left( \frac{v}{v_e} \right)^{C_5}} \right)
\end{equation}

Thus $\psi_1(0.35r_0) \rightarrow > 1$, which gives us (equation \ref{eq:svet2007gain}):

\begin{equation}
    m_\text{atmgain} = y_\text{imp} m_\text{imp} \left( 1 - \left(\psi_1(0.35r_0)\uparrow \frac{M_\text{aE}}{M_a} \left( C_7 \frac{v_e}{v} \right)^{0.25} +\zeta_v(v) \frac{2}{\pi} \arctan\left(\zeta_v(v)^4 \left( \frac{r_0}{H} \right)^{1.2}\left( \frac{\rho_m}{\rho_0}\right)^{0.5}\right)\right)\right)
\end{equation}

\begin{figure}
    \centering
    \includegraphics[width=0.8\textwidth,height=\textheight,keepaspectratio]{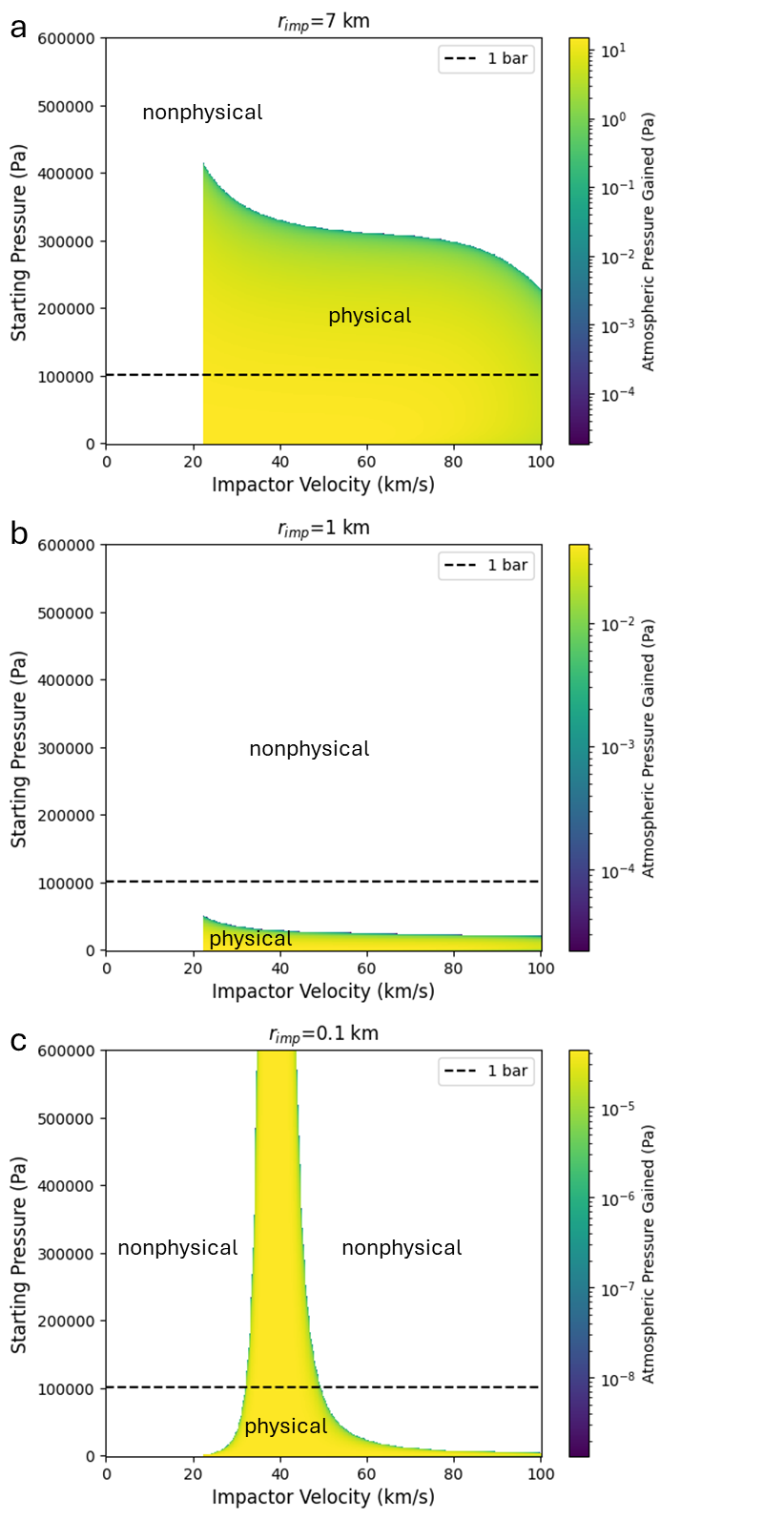}
    \caption{Plot of atmospheric pressure gain from Svetsov 2007's model for varying impactor velocities and starting pressures for present-day Earth parameters.}
    \label{fig:svet2007prob}
\end{figure}

Thus, terms subtracted from unity in the parentheses can become larger than 1 in certain parts of the parameter space (impactor radius, starting atmospheric pressure, and impactor velocity, see figure \ref{fig:svet2007prob}), which results in an overall negative atmospheric gain value.

We also force Svetsov 2007's gains to be $<10^{30}$ kg. This should not occur unless complex numbers crop up, and it is just a failsafe. Complex numbers can occur in equation \ref{eq:svet2007zetav} of the algorithm. If $\frac{v}{v_e}-C_7<0$, then raising it to the power of $C_8$ (which =1.55 for comets and =1.4 for asteroids) can create a complex number, which is nonphysical. If the final value for Svetsov 2007's atmospheric gain is $<0$, $>10^{30} kg$, or a NaN value in the code (complex numbers involved), we set it equal to 0. This process results in the varied shapes of physical values in the parameter space shown in figure \ref{fig:svet2007prob}.

\subsection{Shuvalov 2009}
$\xi$ (equation \ref{eq:shuxi}) can be negative if $v_\text{esc}>v_\text{imp}$. If we were to account for drag, which we do not, the impactor could come in at less than the escape velocity, causing a negative $\xi$ value. However, we need to take the log of $\xi$ later in the algorithm (equations \ref{eq:chiimp} and \ref{eq:chia}). Thus, if $\xi$ is negative, we set it equal to 0.

$\xi$ can also go to infinity if $\rho_0<<\rho_\text{imp}$. In this small atmosphere case, the atmospheric mass loss can go to infinity. $\chi_a(\xi)$ (equation 26) initially peaks around $\xi\approx340$, then drops until a minimum at $\xi\approx4.506\times10^{10}$ where $\chi_a\approx 3.357\times10^{-10}$. $\chi_a(\xi)$ then grows and goes to infinity. Thus, if $\xi>4.506\times10^{10}$, we set $\chi_a= 3.357\times10^{-10}$.

\subsection{de Niem 2012}
If $\frac{v_\text{esc}}{v_\infty}>1$, we set it equal to 1. We also prevent negative values for atmospheric gain or loss. Note that the fixes for the Svetsov 2000 model also apply here, as the $\eta$ value is pulled from that model if $r_\text{imp}<\frac{D_\text{lim}}{2}$.

\subsection{Schlichting 2015}
We made no alterations to this model.

\subsection{Kegerreis 2020}
We prevent the atmospheric mass loss from exceeding the total existing mass of the atmosphere or becoming negative.

\subsection{Composite Model}
If a gain or loss value produced by an individual model required modification, we instead set it to 0. Thus, the composite model only uses the individual models in the portions of the parameter space that do not require modification.

%
%

%

%

\section{Code Details}
\label{app:code} 

\begin{figure}
\noindent\includegraphics[width=\textwidth]{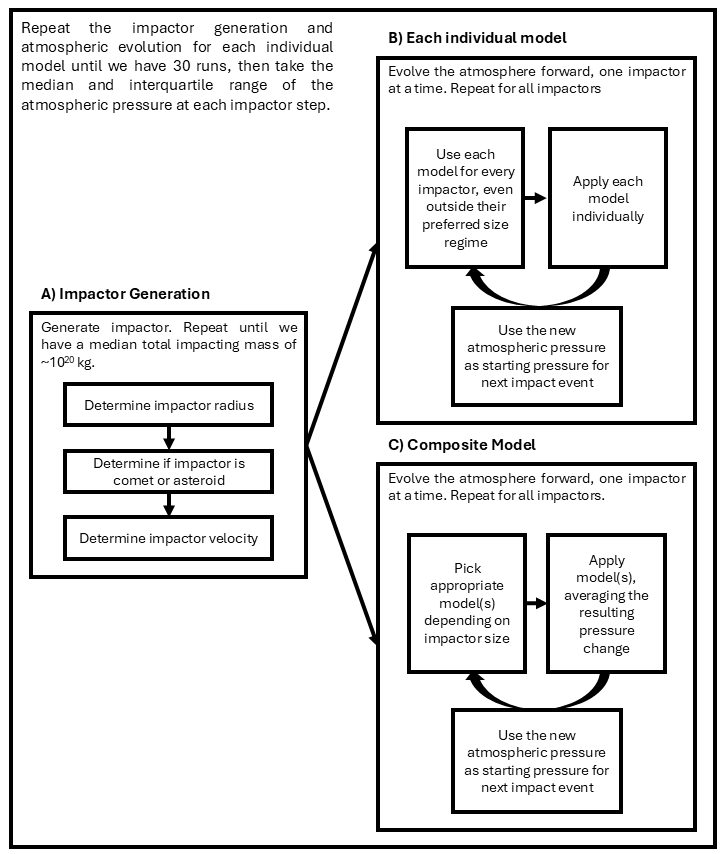}
\caption{We generate an impactor population (Box A), then evolve the atmosphere forward one impactor at a time. We evolve the atmosphere forward using the individual models from the literature, even when this extends beyond the models’ originally intended size ranges (Box B). We also now use our composite model, which only uses the component models in their preferred size ranges, averaging together relevant models (box C).}
\label{fig:flowchartcomp}
\end{figure}
Figure \ref{fig:flowchartcomp} shows the flow of our new methods, with both steps: impactor generation and atmospheric evolution.

\section{Open Research}
Version 23 of Atmospheres and Impacts used for producing data and generating plots included above is preserved at https://github.com/mikaylahuffman/Atmospheres-and-Impacts, publicly available for download \cite{atmospheresandimpactscode}.






\acknowledgments
We thank the two anonymous reviewers for their useful comments that improved the quality of this manuscript. This work was funded by NASA DRIVE grant number 80NSSC20K0594 and NASA ICAR grant number 80NSSC23K1358. This work utilized the Alpine High-Performance Computing resource at the University of Colorado Boulder. Alpine is jointly funded by the University of Colorado Boulder, the University of Colorado Anschutz, Colorado State University, and the National Science Foundation (award 2201538).



%
\bibliography{biblio} 




%
%
%
%
%

\end{document}